\pdfoutput=1
\documentclass[cernpreprint,english]{na61doc}

\usepackage{lineno}
\usepackage{enumerate}
\usepackage{todonotes}

\usepackage{colortbl}
\definecolor{darkred}{rgb}{0.5,0,0}
\definecolor{darkblue}{rgb}{0,0,0.5}
\definecolor{firebrick}{rgb}{0.75,0.125,0.125}
\definecolor{darkgreen}{rgb}{0,0.5,0}
\usepackage{tabularx}
\usepackage{lineno}
\usepackage{cite}
\usepackage[label font=bf,labelformat=simple]{subfig}
\usepackage{caption}

\usepackage[colorlinks=true,linkcolor=firebrick,citecolor=darkgreen,urlcolor=darkblue]{hyperref}

\newcommand{\eV}{\ensuremath{\mbox{e\kern-0.1em V}}\xspace}
\newcommand{\TeV}{\ensuremath{\mbox{Te\kern-0.1em V}}\xspace}
\newcommand{\GeV}{\ensuremath{\mbox{Ge\kern-0.1em V}}\xspace}
\newcommand{\MeV}{\ensuremath{\mbox{Me\kern-0.1em V}}\xspace}
\newcommand{\GeVc}{\ensuremath{\mbox{Ge\kern-0.1em V}\!/\!c}\xspace}
\newcommand{\MeVc}{\ensuremath{\mbox{Me\kern-0.1em V}\!/\!c}\xspace}
\newcommand{\AGeV}{\ensuremath{A\,\mbox{Ge\kern-0.1em V}}\xspace}
\newcommand{\AGeVc}{\ensuremath{A\,\mbox{Ge\kern-0.1em V}\!/\!c}\xspace}

\newcommand{\cm}{\ensuremath{\mbox{cm}}\xspace}

\newcommand{\dd}{\ensuremath{{\textrm d}}\xspace}
\newcommand{\dedx}{\ensuremath{\dd E\!/\!\dd x}\xspace}

\newcommand{\pt}{\ensuremath{p_{T}}\xspace}
\newcommand{\y}{\ensuremath{{y}}\xspace}

\newcommand{\mt}{\ensuremath{m_{\textrm T}}\xspace}
\newcommand{\pp}{\mbox{\textit{p+p}}\xspace}
\newcommand{\pA}{\mbox{\textit{p}+A}\xspace}

\newcommand{\pim}{\ensuremath{\pi^-}\xspace}
\newcommand{\pip}{\ensuremath{\pi^+}\xspace}

\newcommand{\Xim}{\ensuremath{\Xi^{-}}\xspace}
\newcommand{\Xip}{\ensuremath{\overline{\Xi}{^+}}\xspace}

\newcommand{\Urqmd}{{\scshape U}r{\scshape qmd}\xspace}

\newcommand{\Epos}{{\scshape Epos}\xspace}

\newcommand{\Ampt}{{\scshape Ampt}\xspace}
\newcommand{\Hijing}{{\scshape Hijing}\xspace}
\newcommand{\PHSD}{{\scshape Phsd}\xspace}
\newcommand{\SmashModel}{{\scshape Smash}\xspace}

\newcommand{\CernVM}{\textsc{Cern\-\kern-0.05emVM}\xspace}

\graphicspath{plots}

\ShineTitle{Measurements of $\Xim$ and $\Xip$ production in proton-proton interactions at $\sqrt{s_{NN}}$ = 17.3~\GeV \\ in the \NASixtyOne experiment}
\PreprintIdNumber{CERN-EP-2020-102}
\ShineJournal{EPJ C}

\ShineAbstract{ %
The production of $\Xi(1321)^{-}$ and $\overline{\Xi}(1321)^{+}$ hyperons in inelastic \pp interactions is studied in a fixed target experiment at a beam momentum of 158~\GeVc. Double differential distributions in rapidity \y and transverse momentum \pt are obtained from a sample of 33M inelastic events. They allow to extrapolate the spectra to full phase space and to determine the mean multiplicity of both $\Xim$ and $\Xip$. 
The rapidity and transverse momentum spectra are compared to transport model predictions. The 
$\Xim$ mean multiplicity in inelastic \pp interactions at 158~\GeVc is used to quantify the strangeness enhancement in A+A collisions at the same centre-of-mass energy per nucleon pair.
}
\begin{document}
\maketitle

\section{Introduction}
Hyperons are made up of one or more strange valence quarks. In \pp interactions the initial state has no constituent strange quarks. Thus, hyperons are excellent probes of the dynamics of \pp interactions. As a result hyperon production has been studied in a long series of experiments on elementary \pp interactions as well as proton-nucleus and nucleus-nucleus collisions. Nevertheless, the experimental data on hyperon production in \pp interactions are incomplete, and their interpretation is all but conclusive.
At the same time rather impressive efforts have been invested into studies of hyperon production in 
nucleus-nucleus interactions, because strangeness carrying particles are expected to have different characteristics when produced in hadron-hadron and nucleus-nucleus collisions. These differences 
increase with the strangeness content of the particle. Thus hyperons containing two or three strange quarks are especially important.
This subject has been first brought up in connection with the search for the Quark Gluon Plasma, a ”deconfined” state of matter in high energy nucleus-nucleus interactions~\cite{Rafelski:1982pu}. The authors predict an enhanced production of strange particles, especially of doubly strange hyperons. 
In this context "enhanced" means that the multiplicity normalized to the number of nucleons participating in the collision is significantly greater in central A+A than in inelastic \pp collision at the same centre-of-mass energy per nucleon pair ($\sqrt{s_{NN}}$).

In the absence of reliable results on multi-strange hyperon production in inelastic \pp interactions in the SPS energy range, however, such claims are often based on assumptions as e.g. the validity of an elementary reference extracted from hadron-nucleus data. A number of complex nuclear effects enter here which are difficult to control quantitatively. This is
why \NASixtyOne  has embarked upon a systematic study of hyperon production in an experimental programme which covers hadron-proton, hadron-nucleus, and nucleus-nucleus collisions~\cite{Aduszkiewicz:2015dmr, Abgrall:2015hmv, Abgrall:2013wda, Aduszkiewicz:2019vck}. These fixed target measurements employ the same detector and beam momenta from 13 to 158~\GeVc per nucleon.

This publication presents measurements of $\Xim$ and $\Xip$ hyperon production in inelastic \pp interactions at 158~\GeVc corresponding to $\sqrt{s_{NN}}$=17.3~\GeV. A total of 53 million minimum bias events were recorded. 

\section{The \NASixtyOne detector}
\begin{figure}
\centering
\includegraphics[width=0.9\textwidth]{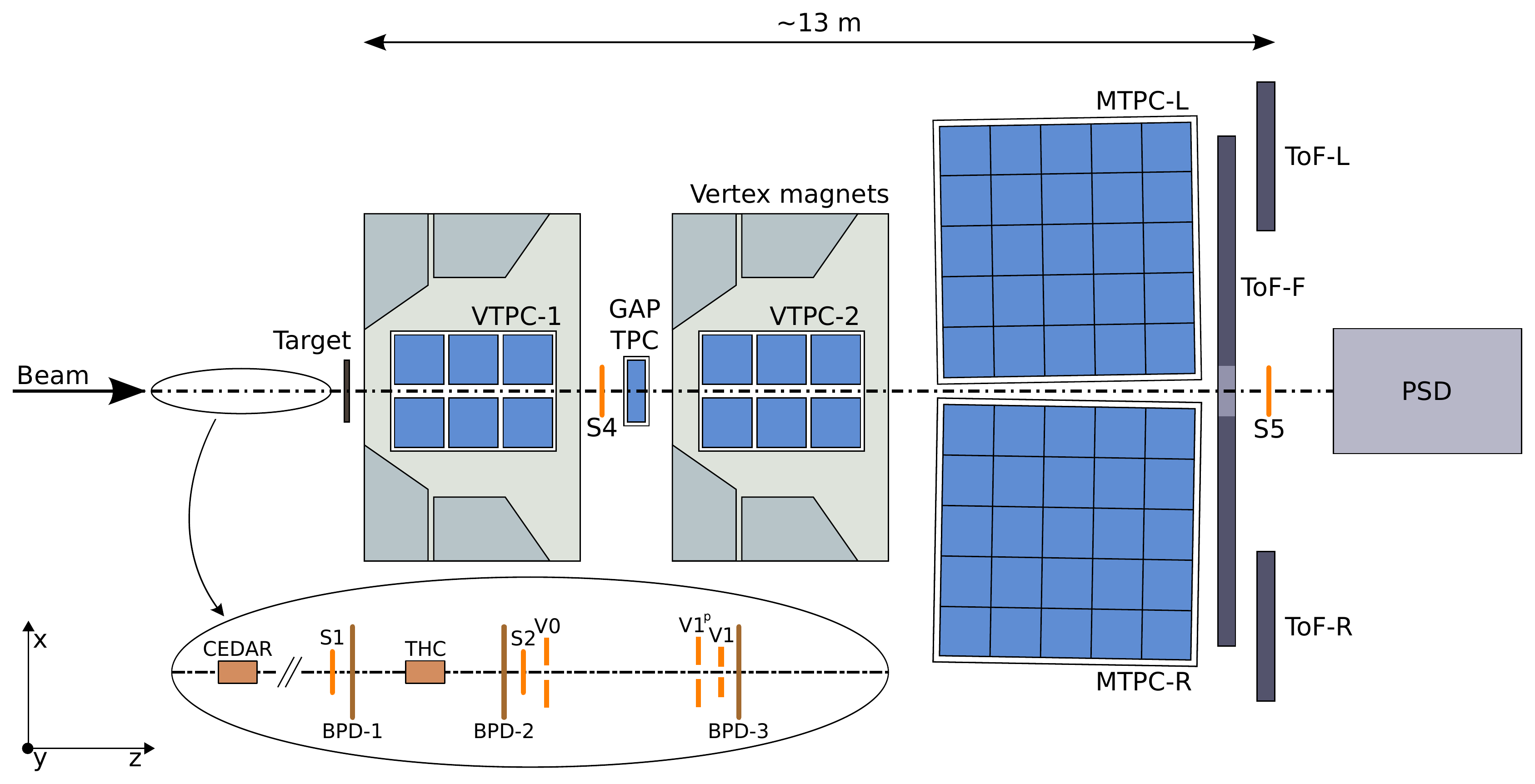}
\caption{(Color online) Schematic layout of the NA61/SHINE experiment
at the CERN SPS (horizontal cut, not to scale).
The orientation of the \NASixtyOne coordinate system is shown on the picture. 
The nominal beam direction is along the $z$ axis. The magnetic field bends charged particle
trajectories in the $x$-$z$ plane. The electron drift direction in the TPCs is along  the $y$ (vertical) axis. 
}
\label{fig:detector-setup}
\end{figure}

\NASixtyOne is a fixed target experiment employing a large acceptance hadron spectrometer situated in the North Area H2 beam-line of the CERN SPS~\cite{Abgrall:2014fa}. A schematic layout is shown in Fig.~\ref{fig:detector-setup}. 
The main components of the detection system are four large volume Time Projection Chambers (TPC). Two of them, called Vertex TPCs (VTPC-1, VTPC-2), are located downstream of the target inside superconducting magnets with combined  maximum bending power of 9~Tm. The MTPCs and two walls of pixel Time-of-Flight (ToF-L/R) detectors are placed symmetrically to the beamline downstream of the magnets. A GAP TPC between VTPC-1 and VTPC-2 improves the acceptance for high-momentum forward-going tracks.

A secondary beam of positively charged hadrons at a momentum of 158~\GeVc was used to collect the data for the analysis presented in this paper. 
This beam was produced by 400~\GeVc protons on a Be-target. The primary protons were extracted from the SPS in a slow extraction mode with a flat-top of 10 seconds. Protons produced together with other particles in the Be-target constitute the secondary hadron beam. The former are identified by two Cherenkov counters, a CEDAR~\cite{CEDAR} (either CEDAR-W or CEDAR-N) and a threshold counter (THC). A selection based on signals from the Cherenkov counters allowed to identify the protons in the beam with a purity of about 99\%~\cite{Abgrall:2013qoa}. The beam momentum and intensity was adjusted by proper settings of the H2 beamline magnets and collimators. The current settings in the bending magnets have a precision of approximately 0.5\%. 
Individual beam particles are detected by a set of scintillation counters. Their trajectories are precisely measured by three beam position detectors (BPD-1, BPD-2, BPD-3)~\cite{Abgrall:2014fa}.

A cylindrical target vessel of 20.29~\cm length and 3~\cm diameter was situated upstream of the entrance window of VTPC-1 (centre of the target z=-580 cm in the NA61/SHINE coordinate system, where $z$=0 is at the centre of the magnet around VTPC-2). The vessel was filled with liquid hydrogen corresponding to an interaction length of 2.8\%.
The ensemble of vessel and liquid hydrogen constitute the "Liquid Hydrogen Target" (LHT).
Data were taken with full and empty  LHT.

Interactions in the target are selected with the trigger system by requiring an incoming beam proton and no signal from S4, a small 2~\cm diameter scintillation counter placed on the beam trajectory between the two vertex magnets (see Fig.~\ref{fig:detector-setup}). This
minimum bias trigger fires, if no charged particle is detected on the beam trajectory downstream of the target.

\section{Event selection}

Inelastic \pp events were selected using the following criteria:
\begin{enumerate}[(i)]
\item no off-time beam particle detected within a time window of $\pm$2$~\mu$s around the time of the trigger particle,
\item beam particle trajectory measured in at least three planes out of four of BPD-1 and BPD-2 and in both planes of BPD-3,
\item the primary interaction vertex fit converged,
\item $z$ position of the interaction vertex (fitted using the beam trajectory and TPC tracks) not farther away than 20~\cm from the center of the LHT,
\item events with a single, positively charged track with laboratory momentum close to the beam momentum (see Ref.~\cite{Abgrall:2013qoa}) are rejected, which eliminates most of the elastic scattering reactions.
\end{enumerate}

The data sample used in this paper was registered in 2009, 2010 and 2011. After the above selection of inelastic events it is reduced to 33 millions.

\section{$\Xi$ reconstruction method}

Particle trajectories (tracks) were reconstructed using an appropriate selection of TPC-clusters. The corresponding momenta were calculated on the basis of the trajectories and the magnetic field values along the trajectory. Fits provided the momentum vectors at the main interaction vertex and at the first measured point.
The $\Lambda$ (here and in the following the line of arguments holds also for the anti-particles)    candidates are found by pairing tracks with appropriate mass and charge assignments. The corresponding particles are tracked backwards through the \NASixtyOne magnetic field from the
first track point, which is required to lie in one of the VTPCs. This backtracking is performed
in 2~cm steps in the $z$ (beam) direction. At each step the separation in the transverse coordinates $x$ and $y$ is evaluated and the minimum is searched for. A pair is considered a $\Lambda$ candidate if the distances in the $x$ and $y$ directions at the minimum are both below 1~cm. Using the distances at the two neighbouring space points around the found  minimum the point of closest approach is found by interpolation. This point is the first approximation of the $\Lambda$ decay point. Its position together with the momenta of the particles at this point are used as input for a 9 parameter fit using the Levenberg-Marquardt procedure~\cite{press_etal_1992}. It provides the momentum vectors of both decay particles and the final coordinates of the $\Lambda$ decay point.

To find the $\Xi$ candidates, all $\Lambda$ candidates are combined with charged pion tracks of appropriate charge sign. A $\Xi$ candidate fitting procedure with 13 parameters~\cite{press_etal_1992} is applied, using as parameters the decay position of the $\Lambda$ candidate, the momentum vectors of both $\Lambda$ decay particles, the momentum vectors of the daughter particles, and finally the $z$
position of the $\Xi$ decay point. 
The $x$ and $y$ coordinates of the $\Xi$ decay position are not subject of the minimization, as they are calculated using the fit results and momentum conservation. This procedure yields the decay position and the momentum vector of the $\Xi$ candidate.

\section{Selection of $\Xi$ candidates}

Several cuts are applied to track parameters and decay topologies in order to minimize the combinatorial background and to maximize the signal to background ratio. They represent a compromise between the size of the hyperon signal and the signal to background ratios in the various invariant mass distributions (see Section~\ref{sec:signal}).

To ensure good track quality and well defined momenta tracks are accepted only if they have at least 10 clusters in either VTPC-1 or VTPC-2. The identification of charged pions and protons is based on the specific energy loss (\dedx) recorded in the TPCs for the corresponding tracks. The appropriate mass is assigned, if the energy loss is within a $\pm 3~\sigma_{\dedx}$ window around the expectation value given by a Bethe–Bloch parametrization adjusted to the \dedx measurements. 

A rapidity dependent cut is applied on the distance between the primary and the secondary $\Xi$ vertex. Its values are shown in Table~\ref{tab:distance}. Rapidity values in the paper are given in the centre-of-mass frame. Additionally, the decay vertex of the $\Lambda$ is requested to be located downstream in $z$ from the $\Xi$ decay vertex. Also a mass window of $\pm 15~\MeV$ around the nominal PDG value~\cite{PhysRevD.98.030001} is applied in the invariant mass distribution of the $\Lambda$ candidates to improve the selection of the $\Xi$ candidates. The combinatorial background under the $\Xi$ signal in the invariant mass distribution (see Section~\ref{sec:signal}), formed by tracks originating from the main vertex is reduced by applying a cut on the distance of closest approach (DCA) of the $\Xi$ trajectory at the $z$ position of the main vertex. Since the DCA resolution
is approximately twice better in $y$ than in $x$ directions, the cut is implemented as: 
$\sqrt{\left(bx_{\Xi}\right)^{2}+(by_{\Xi}/0.5)^{2}}<$ 1.0~cm. About half of the background is removed by this cut with the signal essentially unchanged. 
The charged pion daughter of the $\Xi$ originates from a displaced vertex. Thus the background is further reduced by requiring that the DCA of the extrapolated daughter track at the $z$ position of the main vertex is 
$\sqrt{\left(bx_{daughter}\right)^{2}+(by_{daughter}/0.5)^{2}}>$~0.5~cm. This reduces the background by about 10\%, while only approximately 2\% of the signal is removed.

\begin{table}[ht]
\caption{$\Xi$ distance cut between primary and secondary vertex in the $z$ (beam) direction for different rapidities.\label{tab:distance}}
\centering
			\begin{tabular}{|c||c|c|c|c|}
			    \multicolumn{5}{c}{}\\
				\hline
				&&&&\\
				$\Xi$ rapidity & $\y < -1.75$ & $-1.75 < \y < 0.75$ & $0.75 < \y < 1.25$ & $1.25 < \y$ \\
				&&&&\\
				\hline
				&&&&\\
				minimum decay length &  0~cm & 5~cm & 12~cm & 20~cm \\
				&&&&\\
				\hline
			\end{tabular}
			
\end{table}

\section{Signal extraction}\label{sec:signal}

\begin{figure*}[ht!]
\centering
\includegraphics[width=.48\textwidth]{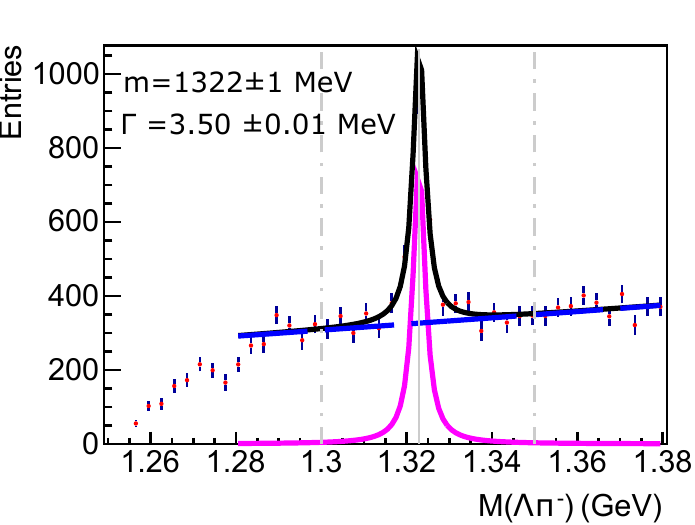}
\includegraphics[width=.48\textwidth]{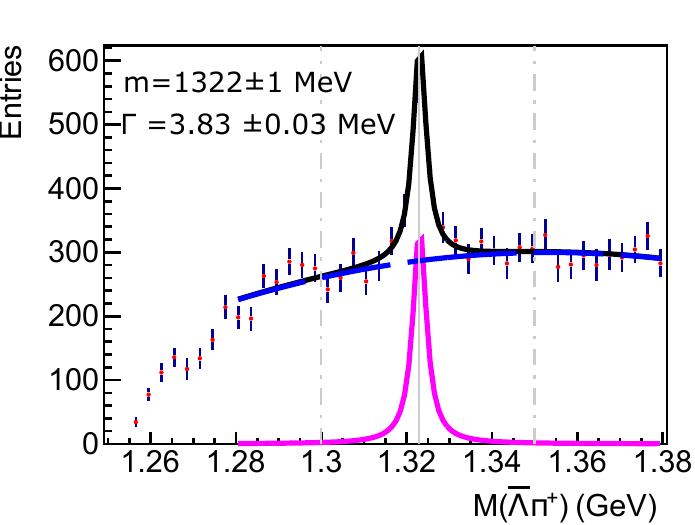}
\caption{(Color online) \textit{Left:} 
The $\Lambda$\pim invariant mass spectrum of
\Xim candidates for rapidity $\y$ between -0.25 and 0.25 and transverse momentum \pt from 0.4 to 0.6~\GeVc. Magenta line represents the fitted Lorentzian function and blue one shows the fitted background, black line represents their sum. The vertical solid gray line shows the nominal PDG $\Xi$ mass, dashed lines show the integration range used. 
\textit{Right:} Analogous  $\overline{\Lambda}$\pip invariant mass spectrum of \Xip candidates.}
\label{fig:Xifit}
\end{figure*}

For each $\Xi$ candidate the invariant mass was calculated assuming the $\Lambda$ and pion masses for the reconstructed candidate daughter particles in suitably selected ($\y$-\pt) bins.
A careful evaluation of the combinatorial background allows to determine the number of $\Xim$  and $\Xip$
in each bin. The corresponding procedure consists of a fit of a signal and background function
to the experimental distribution using $\chi^{2}$ minimization. The signal is described by the Lorentzian function:
\begin{equation}
    \label{eq:lorenz}
    L(m)=\frac{1}{\pi}\frac{\frac{1}2{}\Gamma}{(m-m_{\Xi})^{2}+(\frac{1}{2}\Gamma)^{2}},
\end{equation}
where $m_{\Xi}$ is the center of the distribution and $\Gamma$ is a parameter specifying the width.
The background is parametrized by a 2nd order polynomial (3th and 4th order polynomial for the estimation of the systematic uncertainty - see Section~\ref{sec:systematics}). The fit is performed over the mass range from 1.29 to 1.38~\GeV. It is important to note that the extracted yield varies smoothly, when extending the mass range, and stabilizes beyond the above mentioned mass interval. The parameters describing the background were fixed using this interval. The signal is then determined by subtracting the background function from the experimental invariant mass spectrum. In order to limit the propagation of statistical background fluctuations into the signal, the mass range for this extraction is restricted to the base width of the hyperon mass distribution as given by the Lorentzian function with an additional extension of $\pm 12~\MeV$. Figure~\ref{fig:Xifit} shows the invariant mass distribution of $\Xim$  and $\Xip$ candidates for the central rapidity bin and transverse momenta around 0.5~\GeVc. The black, blue, and magenta lines show the combined, background and signal fit functions, respectively.

The extracted mass ($m_{\Xi} = 1322 \pm 1$~\MeV) of the measured hyperons
match the PDG value ($m_{\Xi} = 1321.71\pm0.07$~\MeV)~\cite{PhysRevD.98.030001}. The fitted widths are close to expectations as given by the analysis of inelastic \pp interactions generated by \Epos~1.99 with full detector simulation and standard track and $\Xi$ reconstruction procedures.

\section{Corrections factors for yield determination}

In order to determine the true numbers of charged hyperons produced in inelastic \pp interactions a 
set of corrections was applied to the extracted raw results. 

The triggered and accepted events comprise interactions with the target vessel and other material in the vicinity of the target. To estimate the fraction of those events about 10\% of the data were collected without the liquid hydrogen in the target vessel. The signal extraction procedure described in Section~\ref{sec:signal} was applied to these events (1.3 millions events was selected), and the resulting suitably normalized yields were subtracted from the results of the analysis of the data sample with full target vessel. This correction was applied for each  ($\y$, \pt) bin.  
The normalization of the empty target data was based on the fitted vertex $z$ distribution. The ratio of the
numbers of events with the fitted vertex outside of the target (in the range from -400~\cm to -200~\cm) was calculated for full and empty target data and used subsequently as the normalization factor~\cite{Aduszkiewicz:2017sei, Abgrall:2013qoa}.

A detailed Monte-Carlo simulation is performed to quantify the losses due to acceptance limitations, detector inefficiencies, reconstruction shortcomings, analysis cuts, and re-interactions in the target. This
simulation used complete events produced by the \Epos~1.99~\cite{Pierog:2009zt} event generator hitting a hydrogen target of appropriate length. The generated particles in each Monte-Carlo event are tracked through the detector using a GEANT3~\cite{Geant3} simulation of the \NASixtyOne apparatus. 
They are then reconstructed with the same software as used for real
events. Numerous variables are confirmed to be similar to data, such as residual distributions, widths of mass
peaks, track multiplicities and their differential distributions, number of events with no tracks in the detector, as well the cut variables and others.
The reconstructed Monte-Carlo events are then analyzed in the same way as the experimental data. 

A correction factor is computed for each ($\y$, \pt) bin:
\begin{equation}
    C_{F} = n_{generated}^{MC}/n_{rec}^{MC},
\end{equation}
where $n_{rec}^{MC}$ is the number of reconstructed, selected and identified $\Xi$s normalized to the number of analyzed events, and $n_{generated}^{MC}$ is the number of $\Xi$s generated by \Epos~1.99 normalized to the number of generated inelastic interactions. The raw multiplicity of extracted particles is multiplied by this correction $C_{F}$ in order to determine the true $\Xim$ and $\Xip$ yields. These correction factors also
include the branching fraction of the decay into the non–measured in \NASixtyOne channels (99.887\% of the $\Xi$ hyperons  decay into registered channels).

The contribution of $\Omega$ decays to the $\Xi$ yield in the final state is neglected. Typically the multiplicity of $\Omega$s is approximately a factor of 10 lower than the $\Xi$ multiplicity (at pp 7~\TeV collisions \cite{Abelev:2012309}). The small branching fraction of $\Omega$ decays into charged $\Xi$s and the small $\Omega$ production probability imply that its contribution is significantly below 1\%.

\begin{figure*}[ht!]
\centering
\includegraphics[width=.60\textwidth]{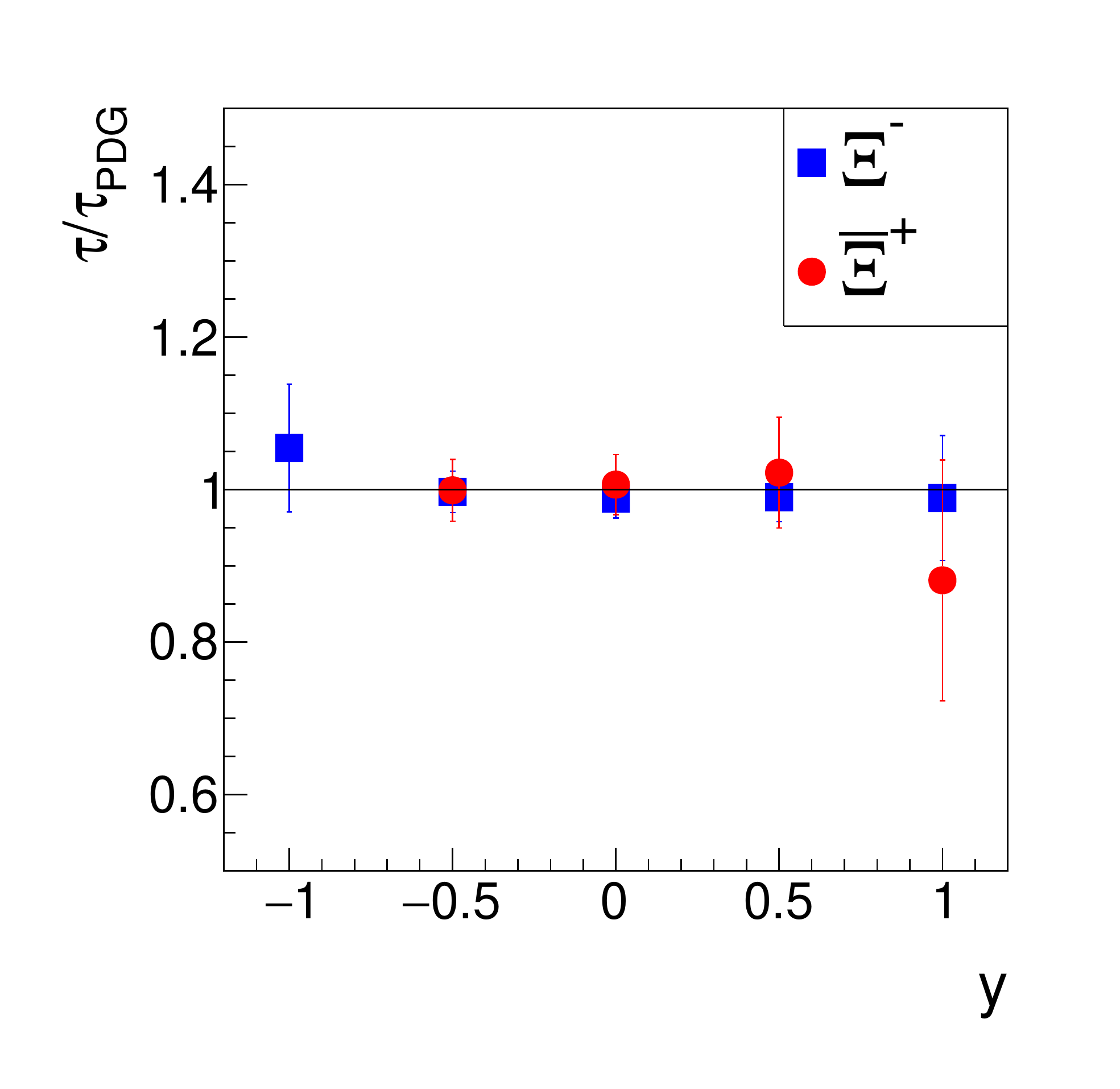}
\caption{(Color online) Measured lifetime ratio $\tau/\tau_{PDG}$ for $\Xim$ (blue squares) and $\Xip$ (red circles) as a function of center of mass rapidity. Only statistical uncertainties are shown.}
\label{fig:tau}
\end{figure*}

Additionally, analysis in rapidity and lifetime bins was performed. Obtained $\Xim$ and $\Xip$ lifetimes are consistent with the PDG ones: $\tau_{PGD}=1.639\times10^{-10}$~s and $\tau_{PGD}=1.700\times10^{-10}$~s for $\Xim$ and $\Xip$, respectively. The resulting $\tau/\tau_{PDG}$ ratio as a function of center of mass rapidity is shown in Fig.~\ref{fig:tau}.

\section{Systematic uncertainties}\label{sec:systematics}

Possible systematic biases of final results (spectra and mean multiplicities)
are due to imperfectness of the Monte Carlo procedure - physics models and detector response simulation - used to calculate the correction factors.

To determine the magnitude of the different sources of possible biases several tests were done:
\begin{enumerate}[(i)]
    \item Methods of event selection.
    
    Not all events  which have tracks stemming from interactions of off-time beam particles are removed. A possible bias due to this effect was estimated by changing by $\pm~1~\mu s$ the width of the time window in which no second beam particle is allowed  with respect to the nominal value of $\pm$2$~\mu$s. The maximum difference of the results was taken as the bias due to the selection. It was estimated to be 2-4\%.
	
	Another source of a possible bias are losses of inelastic events due to the interaction trigger. The S4 detector trigger condition selects mainly inelastic interactions and vetoes elastic scattering events. However, it will miss some of the inelastic events. To estimate the possible loss of $\Xi$s, simulations were done with and without the S4 trigger condition. 
	The difference between these two results was taken as another contribution to the systematic uncertainty. 
	The bias due to the interaction trigger was calculated as the difference between these two results and it is 3-6
	
	
	\item Methods of $\Xim$ and $\Xip$ candidates selection.
	
	To estimate the bias related to the $\Xim$ and $\Xip$ candidate selection the following cut parameters were varied independently: 
	the distance cut between primary and secondary vertex was changed by $\pm$ 1 and $\pm$ 2~cm yielding a possible bias of 2-5\%, 
	the extrapolated impact parameter of $\Xi$s in the $x$ and $y$ direction at the main vertex $z$ position  was changed from $\sqrt{\left(bx_{\Xi}\right)^{2}+(by_{\Xi}/0.5)^{2}}<$ 1.0~cm to $\sqrt{\left(bx_{\Xi}\right)^{2}+(by_{\Xi}/0.5)^{2}}<$ 0.5 and 2~cm yielding a possible bias of up to 10\%,  
	the DCA of the pion ($\Xi$) daughter track to the main vertex was changed from $\sqrt{\left(bx_{daughter}\right)^{2}+(by_{daughter}/0.5)^{2}}>$0.5~cm to 0.2 and 1~cm yielding a possible bias of up to 8\%.

	\item Signal extraction.
	
	The bias due to the signal extraction method were estimated by changing the order of the polynomials used to describe the background from second to third and fourth order yielding an uncertainty of up to 4\%. Varying the invariant mass range used to determine the $\Xi$ yields by a change of $\pm$~12~\MeV with respect to the nominal integration range yielded a possible bias of 2-7\%.

    
    
	\end{enumerate}

The systematic uncertainty was calculated as the square root of the sum of squares of the described possible biases assuming that they are uncorrelated. The uncertainties are estimated for each ($\y$-\pt) bin separately.

\section{Experimental results}\label{seq:results}

This section presents results on inclusive $\Xim$ and $\Xip$ hyperons spectra in inelastic \pp interactions at beam momentum 158~\GeVc. The spectra refer to hyperons produced by strong interaction processes.

\subsection{Spectra and mean multiplicities}
\begin{figure*}[ht!]
\centering
\includegraphics[width=.40\textwidth]{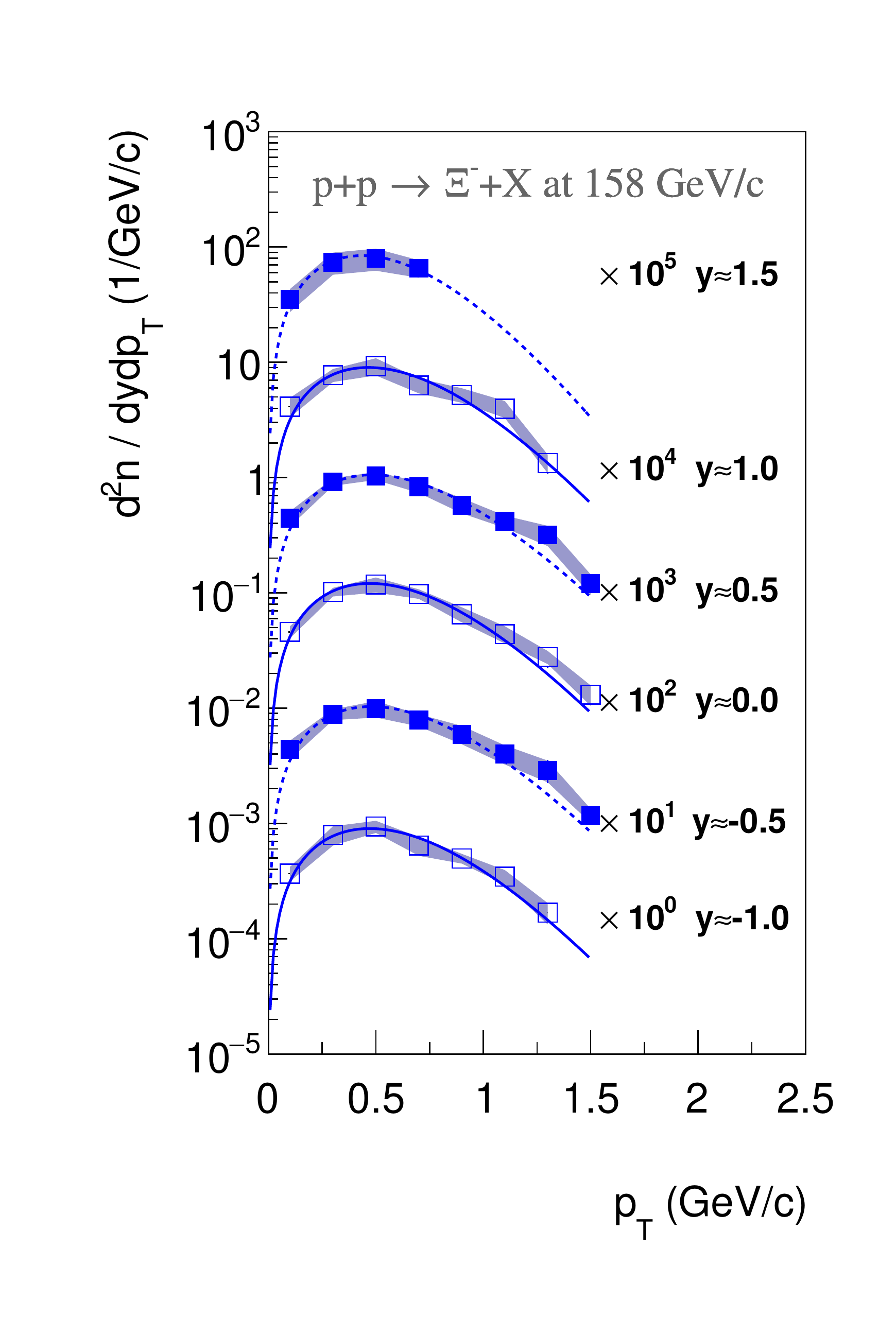}
\includegraphics[width=.40\textwidth]{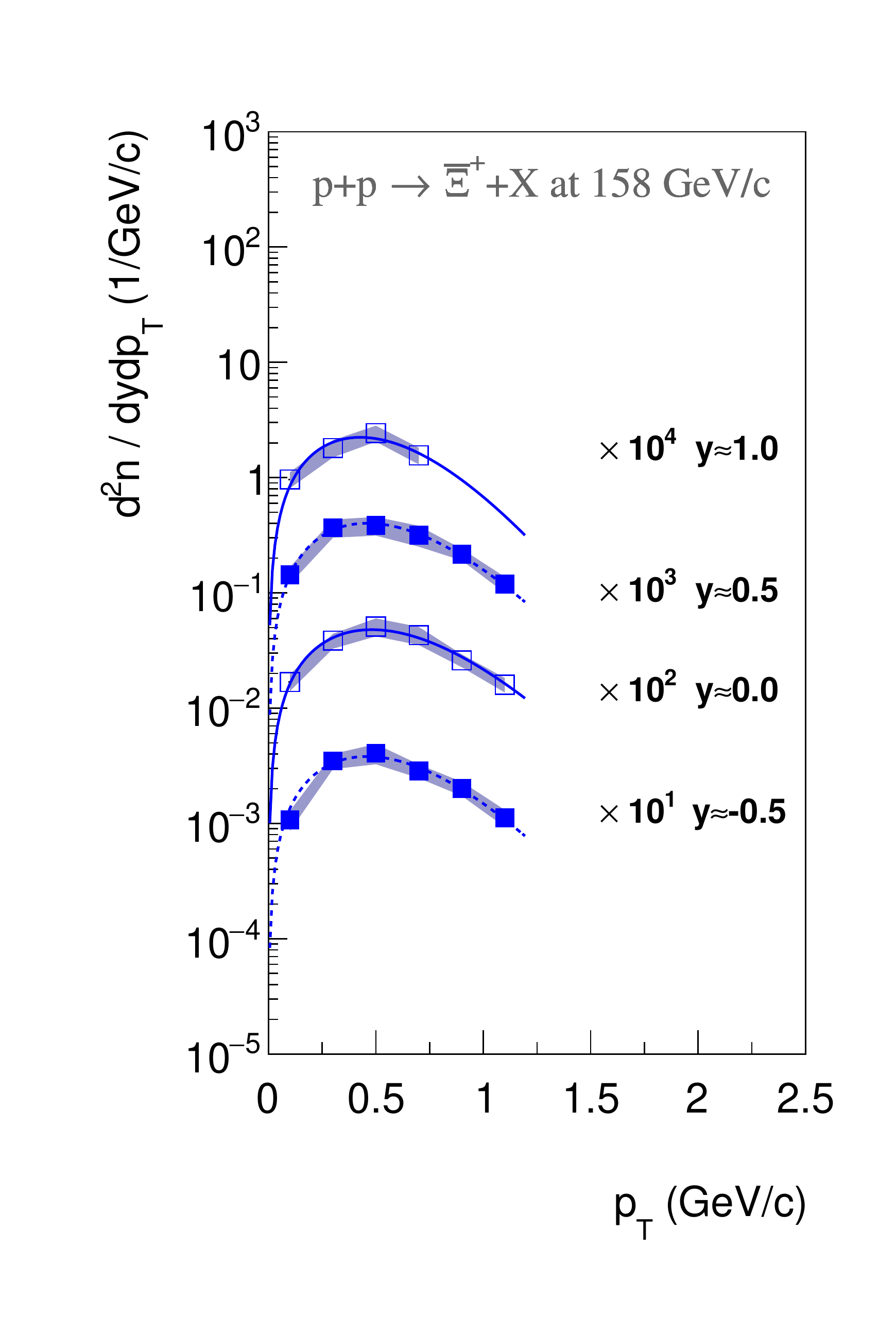}
\caption{(Color online) Transverse momentum spectra in rapidity slices of $\Xim$ (\textit{left}) and $\Xip$ (\textit{right}) produced in inelastic \pp interactions at 158~\GeVc. Rapidity values given in the legends correspond to the middle of the corresponding interval. Statistical uncertainties are smaller than the marker size, shaded bands show systematic uncertainties. Spectra are scaled by the given factors for better separation.}
\label{fig:results}
\end{figure*}

Double differential hyperon yields constitute the basic result of this paper. 
The $\Xim$ ($\Xip$) yields are determined in 6 (4) rapidity and between 4 (4) and 8 (7) transverse momentum bins. The former are 0.5 units and the latter 0.2~\GeVc wide. The resulting ($\y$, \pt) yields are presented in
Fig.~\ref{fig:results} as function of \pt . The transverse momentum spectra can be described by the exponential function~\cite{Hagedorn:1968jf,Broniowski:2004yh}:

\begin{equation}
 \frac{d^{2}n}{d\pt d\y}=\frac{S~c^{2}p_{T}}{T^{2} + m~T}\exp\left(-\frac{\mt - m}{T}\right),
\label{eq:inverse}
\end{equation}

where $m$ is the $\Xi$ mass. The yields $S$ and the inverse slope parameters $T$ are determined by fitting the function to the data points in each rapidity bin. The resulting inverse slope parameters are listed in Table~\ref{tab:dndy}.  The \pt spectra from successive rapidity intervals are scaled by appropriate factors for better visibility. Statistical uncertainties are smaller than the symbol size, shaded bands correspond to systematic uncertainties.
Tables~\ref{tab:results:Xim} and \ref{tab:results:Xip} list the numerical values of the results shown in Fig.~\ref{fig:results}. 

\begin{table}[ht]
\caption{Numerical values of double-differential spectra of $\Xim$ produced in inelastic \pp interactions at 158~\GeVc beam momentum. Rapidity and transverse momentum values correspond to the middle of the presented bin. First value is the particle multiplicity, second represents the statistical uncertainty and third one corresponds to the estimated systematic uncertainty.}\label{tab:results:Xim}
\centering
\small
\begin{tabular}{|c|c|c|c|}
\multicolumn{4}{c}{}\\
\multicolumn{4}{c}{\large{$\Xim$: $\frac{d^{2}n}{d\y d\pt}$ $\times10^{-4}$ (1/\GeVc)}} \\
\multicolumn{4}{c}{}\\
\hline
&&&\\
\pt (\GeVc) & $\y$=-1.0 & $\y$=-0.5 & $\y$=0.0\\
&&&\\
\hline
0.1 & 3.67$\pm$0.35$\pm$0.53 &  4.40$\pm$0.23$\pm$0.82  & 4.59$\pm$0.18$\pm$0.54  \\
0.3 & 7.94$\pm$0.49$\pm$1.51 &  8.82$\pm$031$\pm$0.94   & 10.20$\pm$0.31$\pm$0.87 \\
0.5 & 9.41$\pm$0.55$\pm$1.13 &  9.90$\pm$0.36$\pm$1.60  & 11.8$\pm$0.35$\pm$1.77  \\
0.7 & 6.43$\pm$0.41$\pm$1.19 &  7.90$\pm$0.36$\pm$0.95  & 9.8$\pm$0.38$\pm$0.93   \\
0.9 & 4.96$\pm$0.35$\pm$0.48 &  5.94$\pm$0.36$\pm$1.08  & 6.52$\pm$0.30$\pm$0.97  \\
1.1 & 3.45$\pm$0.38$\pm$0.49 &  4.00$\pm$0.29$\pm$0.75  & 4.39$\pm$0.33$\pm$0.74  \\
1.3 & 1.68$\pm$0.28$\pm$0.35 &  2.89$\pm$0.66$\pm$0.62  & 2.77$\pm$0.54$\pm$0.36  \\
1.5 & -                      &  1.17$\pm$0.14$\pm$0.16  & 1.31$\pm$0.22$\pm$0.26  \\
\hline 
&&&\\
\pt (\GeVc) & $\y$=0.5 & $\y$=1.0 & $\y$=1.5\\
&&&\\
\hline
0.1 & 4.43$\pm$0.24$\pm$0.66    & 4.12$\pm$0.34$\pm$0.79 	 &  3.52$\pm$0.60$\pm$0.79 \\
0.3 &  9.18$\pm$0.31$\pm$0.69   & 7.75$\pm$0.42$\pm$1.00   &  7.35$\pm$0.82$\pm$1.60    \\
0.5 & 10.3$\pm$0.33$\pm$0.82    & 9.27$\pm$0.48$\pm$1.58 	 &  7.95$\pm$0.99$\pm$1.72 \\
0.7 &  8.3$\pm$0.30$\pm$0.94    & 6.31$\pm$0.35$\pm$0.93 	 &  6.55$\pm$0.94$\pm$1.14 \\
0.9 & 5.76$\pm$0.29$\pm$0.87    & 5.20$\pm$0.41$\pm$0.75	 & - \\
1.1 &  4.17$\pm$0.26$\pm$0.51   & 3.96$\pm$0.44$\pm$0.66 	 & - \\
1.3 &  3.18$\pm$0.34$\pm$0.63   & 1.34$\pm$0.25$\pm$0.24	 & - \\
1.5 &  1.20$\pm$0.17$\pm$0.24    & - 									         & - \\
\hline
\end{tabular}
\end{table}

\begin{table}[ht]
\caption{Numerical values of double-differential spectra of $\Xip$ produced in inelastic \pp interactions at 158~\GeVc beam momentum. Rapidity and transverse momentum values correspond to the middle of the presented bin. First value is the particle multiplicity, second represents the statistical uncertainty and third one corresponds to the estimated systematic uncertainty.}\label{tab:results:Xip}
\centering
\small
\begin{tabular}{|c|c|c|c|c|}
\multicolumn{5}{c}{}\\
\multicolumn{5}{c}{\large{$\Xip$: $\frac{d^{2}n}{d\y d\pt}$ $\times10^{-4}$ (1/\GeVc)}} \\ 
\multicolumn{5}{c}{}\\
\hline
&&&&\\
\pt (\GeVc) & $\y$=-0.5 & $\y$=0.0 & $\y$=0.5 & $\y$=1.0\\
&&&&\\
\hline
0.1 & 1.07$\pm$0.14$\pm$0.21 &  1.69$\pm$0.14$\pm$0.28  & 1.44$\pm$0.12$\pm$0.24  &  0.96$\pm$0.15$\pm$0.15  \\
0.3 & 3.49$\pm$0.16$\pm$0.52 &  3.85$\pm$0.16$\pm$0.57  & 3.67$\pm$0.23$\pm$0.66  &  1.80$\pm$0.19$\pm$0.31  \\
0.5 & 4.07$\pm$0.28$\pm$0.81 &  5.10$\pm$0.23$\pm$0.91  & 3.85$\pm$0.23$\pm$0.71  &  2.43$\pm$0.21$\pm$0.39  \\
0.7 & 2.86$\pm$0.18$\pm$0.39 &  4.29$\pm$0.24$\pm$0.78  & 3.17$\pm$0.18$\pm$0.65  &  1.56$\pm$0.16$\pm$0.26  \\
0.9 & 2.01$\pm$0.21$\pm$0.28 &  2.60$\pm$0.20$\pm$0.33  & 2.16$\pm$0.18$\pm$0.25  & -    \\
1.1 & 1.12$\pm$0.15$\pm$0.20 &  1.60$\pm$0.18$\pm$0.23  & 1.19$\pm$0.15$\pm$0.19  & -    \\

\hline
\end{tabular}
\end{table}

\begin{figure}[!ht]
                \begin{center}
                \includegraphics[width=0.6\textwidth]{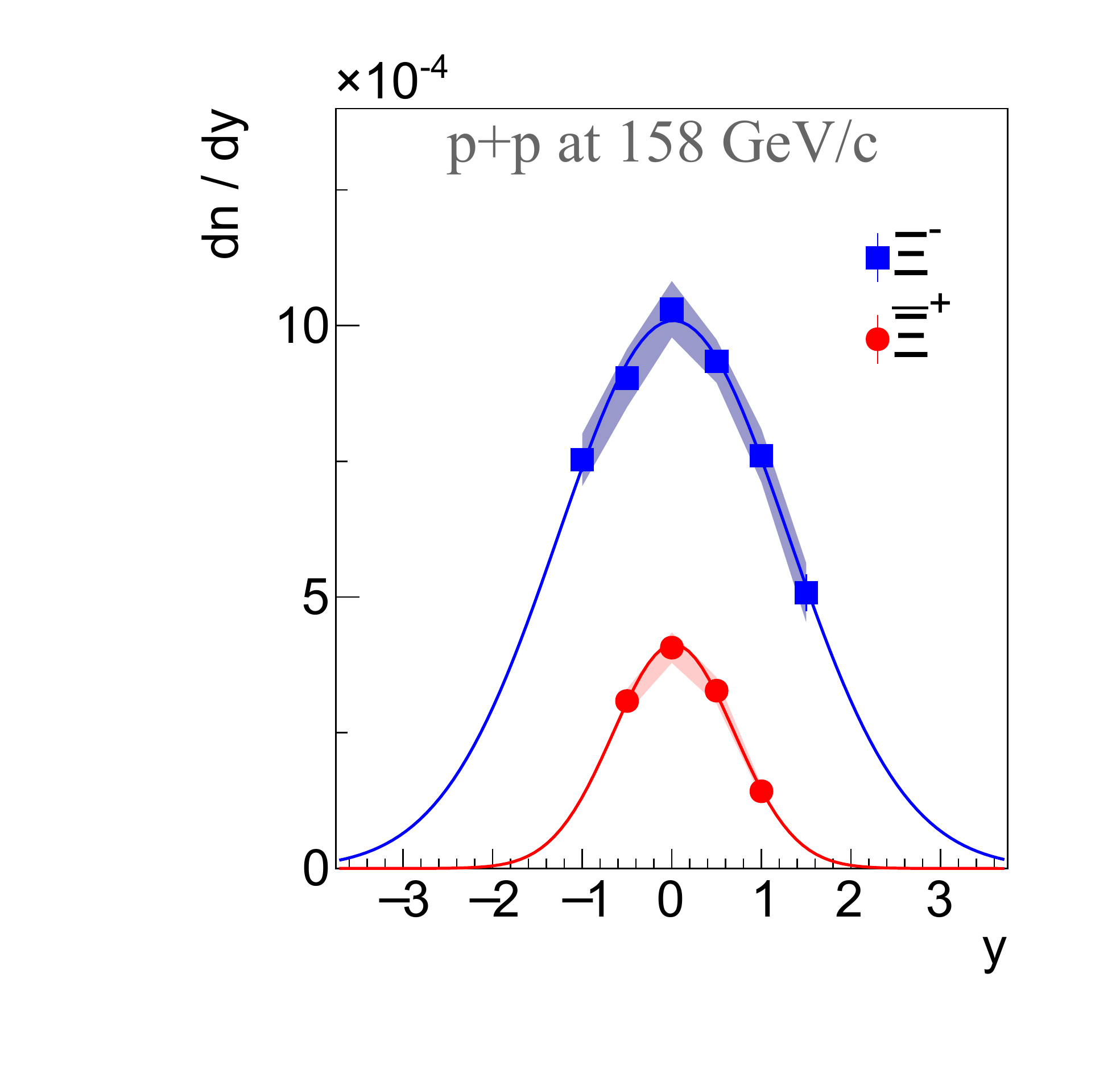}
                \end{center}
                \caption{(Color online) Rapidity spectra of $\Xim$ (blue squares) and $\Xip$ (red circles) produced in inelastic \pp interactions at 158~\GeVc. Statistical uncertainties are smaller than the marker size, shaded bands correspond to systematic uncertainties of the measurements. Curves depict Gaussian fits used to determine total mean multiplicities.}
                \label{fig:finaldndy}
        \end{figure}

Rapidity distributions were then obtained by summing the measured transverse momentum spectra and extrapolating them into the unmeasured regions using the fitted functions given by Eq.~\ref{eq:inverse}.
The resulting rapidity distributions are shown in Fig.~\ref{fig:finaldndy}. The statistical uncertainties are smaller than the symbol size. They were calculated as the square root of the sum of the squares of the statistical uncertainties of the contributing bins. The systematic uncertainties (shaded bands) were calculated
as square root of squares of systematic uncertainty as described in Sec.~\ref{sec:systematics} and  half of the extrapolated yield. The numerical values of rapidity yields and their errors are listed in Table~\ref{tab:dndy}.

\begin{table}[ht]
\caption{Numerical values of rapidity spectra of $\Xim$ and $\Xip$ produced in inelastic \pp interactions at 158~\GeVc beam momentum and fitted inverse slope parameter T (see eq.~\ref{eq:inverse}). Rapidity values correspond to the middle of the presented bin. First value is the multiplicity, second represents the statistical uncertainty and third one corresponds to the estimated systematic uncertainty.}\label{tab:dndy}
\centering
\small
\begin{tabular}{|c|c|c||c|c|}
\multicolumn{5}{c}{}\\
\hline
&&&&\\
$\y$   &   $\Xim$: $\frac{dn}{d\y}$ $\times10^{-4}$ &  $\Xim$: T (\MeV)               &   $\Xip$:  $\frac{dn}{d\y}$ $\times10^{-4}$ &  $\Xip$: T (\MeV)\\
&&&&\\
\hline
-1.0 & 7.53 $\pm$ 0.22 $\pm$ 0.48 & 159 $\pm$ 6 $\pm$ 11 & - & -\\
-0.5 & 9.19 $\pm$ 0.21 $\pm$ 0.53 & 168 $\pm$ 5 $\pm$ 10 & 3.08 $\pm$ 0.09 $\pm$ 0.22 & 150 $\pm$ 7 $\pm$ 10\\
0.0  & 10.3 $\pm$ 0.20 $\pm$ 0.52 & 162 $\pm$ 4 $\pm$ 6 & 4.07 $\pm$ 0.10 $\pm$ 0.28 & 146 $\pm$ 4 $\pm$ 7\\
0.5  & 9.34 $\pm$ 0.16 $\pm$ 0.40 & 169 $\pm$ 4 $\pm$ 8 & 3.27 $\pm$ 0.09 $\pm$ 0.25 & 134 $\pm$ 4 $\pm$ 8\\
1.0  & 7.60 $\pm$ 0.21 $\pm$ 0.49 & 154 $\pm$ 6 $\pm$ 11 & 1.42 $\pm$ 0.07 $\pm$ 0.12 & 135 $\pm$ 15 $\pm$ 16\\
1.5  & 5.08 $\pm$ 0.34 $\pm$ 0.55 & 136 $\pm$ 22 $\pm$ 15 & - & - \\
\hline
\end{tabular}
\end{table}

Gaussian functions were fitted to the rapidity distributions and used to extrapolate into the unmeasured regions. The extrapolation factors for $\Xip$ and $\Xim$ are 1.24 and 1.33, respectively.
Summing the data points and the extrapolated yield add up to the mean multiplicities $\langle \Xim \rangle =$ (3.3 $\pm$ 0.1 $\pm$ 0.6)$\times10^{-3}$ and $\langle \Xip \rangle = $ (7.9 $\pm$ 0.2 $\pm$ 1.0)$\times10^{-4}$. 
The Gaussian function used to determine the multiplicity is a rather arbitrary choice. To study the uncertainty introduced by this choice the same extrapolation factors were computed for the events generated by the two models mentioned in Sec.~\ref{sec:CompModel}. The extrapolation factors obtained from the two models differ by only 5\% and their shapes agree within uncertainties with the one of the experimental data. 
Thus the already assigned systematic error of 50\% of the extrapolated yield is large compared to the uncertainty due to the function used for extrapolation, and no additional uncertainty was added.

\begin{figure*}[ht!]
\centering
\includegraphics[width=.6\textwidth]{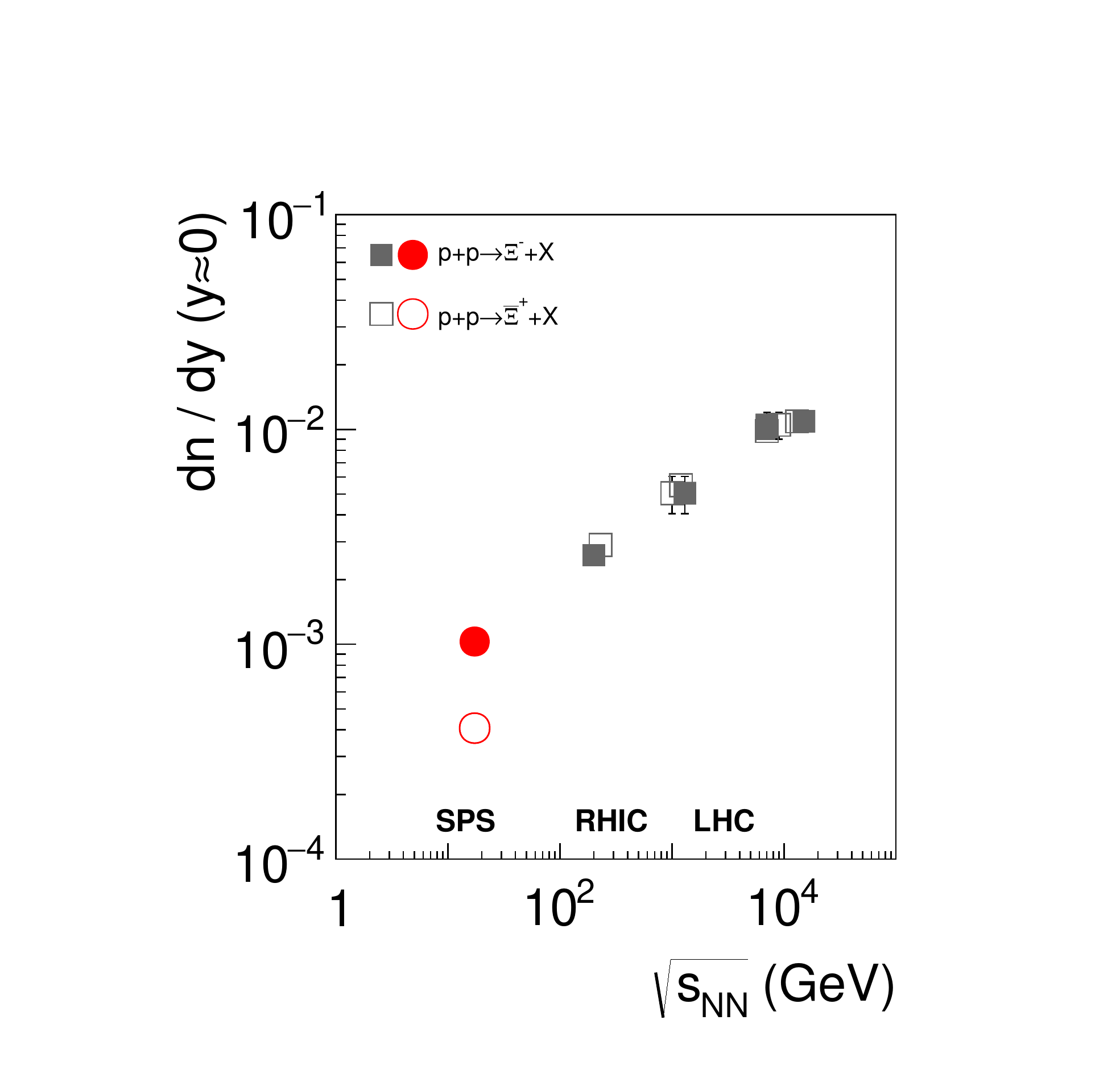}
\caption{(Color online) Mid-rapidity densities (dn/d\y) of $\Xim$ (full symbols) and $\Xip$ (open symbols)
measured in inelastic \pp interactions as a function of centre-of-mass energy $\sqrt{s_{NN}}$. The systematic and statistical uncertainties are smaller than the symbol size. The data are compared to results from STAR at the BNL RHIC measured at $\sqrt{s_{NN}}$ = 200~\GeV~\cite{Abelev:2006cs}, from ALICE at CERN LHC measured at $\sqrt{s_{NN}}$ = 0.9, 7 and 13~\TeV~\cite{Aamodt:2011zza, Abelev:2012jp, Acharya:2019kyh}  and from CMS at the CERN LHC measured at $\sqrt{s_{NN}}$ = 0.9 and 7~\TeV~\cite{Khachatryan:2011tm}.
}
\label{fig:snn}
\end{figure*}

 In Fig.~\ref{fig:snn} we compare the rapidity densities ($dn/d\y$) at mid-rapidity of $\Xim$ and $\Xip$ in inelastic \pp interactions at $\sqrt{s_{NN}}=17.3$~\GeV collisions with results from STAR at the BNL RHIC at $\sqrt{s_{NN}}$ = 200~\GeV~\cite{Abelev:2006cs}, from ALICE at CERN LHC measured at $\sqrt{s_{NN}}$ = 0.9, 7 and 13~\TeV~\cite{Aamodt:2011zza,Abelev:2012jp,Acharya:2019kyh} and from CMS at the CERN LHC measured at $\sqrt{s_{NN}}$ = 0.9 and 7~\TeV~\cite{Khachatryan:2011tm}.
 The yields increase with collision energy by more than an order of magnitude. At 17.3~\GeV the mid-rapidity $\Xip$ yield is almost two times smaller than $\Xim$ yield. This difference vanishes already in the STAR data at 200~\GeV and is negligible beyond.

\subsection{Anti–baryon/baryon ratios}
The production ratio of doubly strange anti-hyperons and hyperons is of special interest since simple string models predict values close to unity because both $\Xim$ and $\Xip$ stem from the pair production process.
The double differential data presented in the previous subsection are therefore presented in the form of ratios (systematic errors were calculated with the procedure of Sec.~\ref{sec:systematics} and not from the systematic uncertainties of the yields which may be correlated). 
The $\Xim/\Xip$ ratios as function of rapidity and transverse momentum are listed in Table~\ref{tab:resultsratio}. The ratio of the rapidity spectra are listed in Table~\ref{tab:dndyratio} and drawn in Fig.~\ref{fig:dnmodel}(c).  We observe little variation with a tendency for a weak maximum around 400~\MeVc in \pt and $\y$=0 in rapidity. The small value of the  ratio of mean multiplicities $\left\langle \Xip \right\rangle / \left\langle \Xim \right\rangle = $  0.24 $\pm$ 0.01 $\pm$ 0.05 emphasizes the strong suppression of $\Xip$ production.

\begin{table}[ht]
\caption{The $\Xip/\Xim$ ratio in inelastic \pp interactions at 158~\GeVc beam momentum. Rapidity and transverse momentum values correspond to the middle of the presented bin. First value is the particle ratio, second represents the statistical uncertainty and third one corresponds to the estimated systematic uncertainty.}\label{tab:resultsratio}
\centering
\small
\begin{tabular}{|c|c|c|c|c|}
\multicolumn{5}{c}{}\\
\multicolumn{5}{c}{\large{$\Xip/\Xim$}} \\ 
\multicolumn{5}{c}{} \\ 
\hline
&&&&\\
\pt (\GeVc) & $\y$=-0.5 & $\y$=0.0 & $\y$=0.5 & $\y$=1.0\\
&&&&\\
\hline
0.1 & 0.243$\pm$0.035$\pm$0.038 &  0.368$\pm$0.034$\pm$0.068  & 0.324$\pm$0.033$\pm$0.073 &  0.233$\pm$0.040$\pm$0.041  \\
0.3 & 0.395$\pm$0.023$\pm$0.058 &  0.378$\pm$0.020$\pm$0.060  & 0.400$\pm$0.028$\pm$0.052 &  0.233$\pm$0.027$\pm$0.038  \\
0.5 & 0.411$\pm$0.032$\pm$0.077 &  0.378$\pm$0.020$\pm$0.060  & 0.372$\pm$0.025$\pm$0.049 &  0.262$\pm$0.026$\pm$0.057  \\
0.7 & 0.362$\pm$0.028$\pm$0.043 &  0.432$\pm$0.023$\pm$0.060  & 0.381$\pm$0.026$\pm$0.070 &  0.247$\pm$0.029$\pm$0.040  \\
0.9 & 0.339$\pm$0.041$\pm$0.045 &  0.438$\pm$0.029$\pm$0.066  & 0.374$\pm$0.036$\pm$0.047 & -    \\
1.1 & 0.280$\pm$0.041$\pm$0.042 &  0.364$\pm$0.049$\pm$0.053  & 0.285$\pm$0.040$\pm$0.050 & -    \\

\hline
\end{tabular}
\end{table}

\begin{table}[ht]
\caption{Ratio of \pt integrated yields versus rapidity of $\Xip$ and $\Xim$ produced in inelastic \pp interactions at 158~\GeVc beam momentum. Rapidity values correspond to the middle of the presented bin. First value is the ratio, second represents the statistical uncertainty and third one corresponds to the estimated systematic uncertainty.}\label{tab:dndyratio}
\centering
\small
\begin{tabular}{|c|c|}
\multicolumn{2}{c}{}\\
\hline
&\\
$\y$   &   $\Xip/\Xim$ \\
&\\
\hline
-0.5 & 0.341 $\pm$ 0.013 $\pm$ 0.026  \\
0.0  & 0.395 $\pm$ 0.012 $\pm$ 0.030 \\
0.5  & 0.350 $\pm$ 0.011 $\pm$ 0.029 \\
1.0  & 0.187 $\pm$ 0.011 $\pm$ 0.018 \\
\hline
\end{tabular}
\end{table}


\subsection{Enhancement factors}
The predicted enhancement of strangeness production in nucleus-nucleus collisions (per participating nucleon) relative to proton-proton reactions was established experimentally 30 years ago \cite{Bartke:1990cn, Alber:1994tz}. It was also found that this enhancement is increasing with the strangeness content of the studied particle\cite{Antinori:1999rw, Antinori:2006ij}. This subsection discusses the system size dependence of the strangeness enhancement in A+A collisions. 
The strangeness enhancement factor $E$ for a given particle species is defined as:
\begin{equation}
 E=\frac{2}{\left<N_{W}\right>}\frac{dn/d\y\left(A+A\right)}{dn/d\y\left(\pp\right)},
\label{eq:enhancment}
\end{equation}

where $\left<N_{W}\right>$ is the number of wounded nucleons in the collision~\cite{Bialas:1976ed}. At SPS energies and above the number of wounded nucleons is close to or equal to the number of participating nucleons.

\begin{figure*}[ht!]
\centering
\includegraphics[width=.49\textwidth]{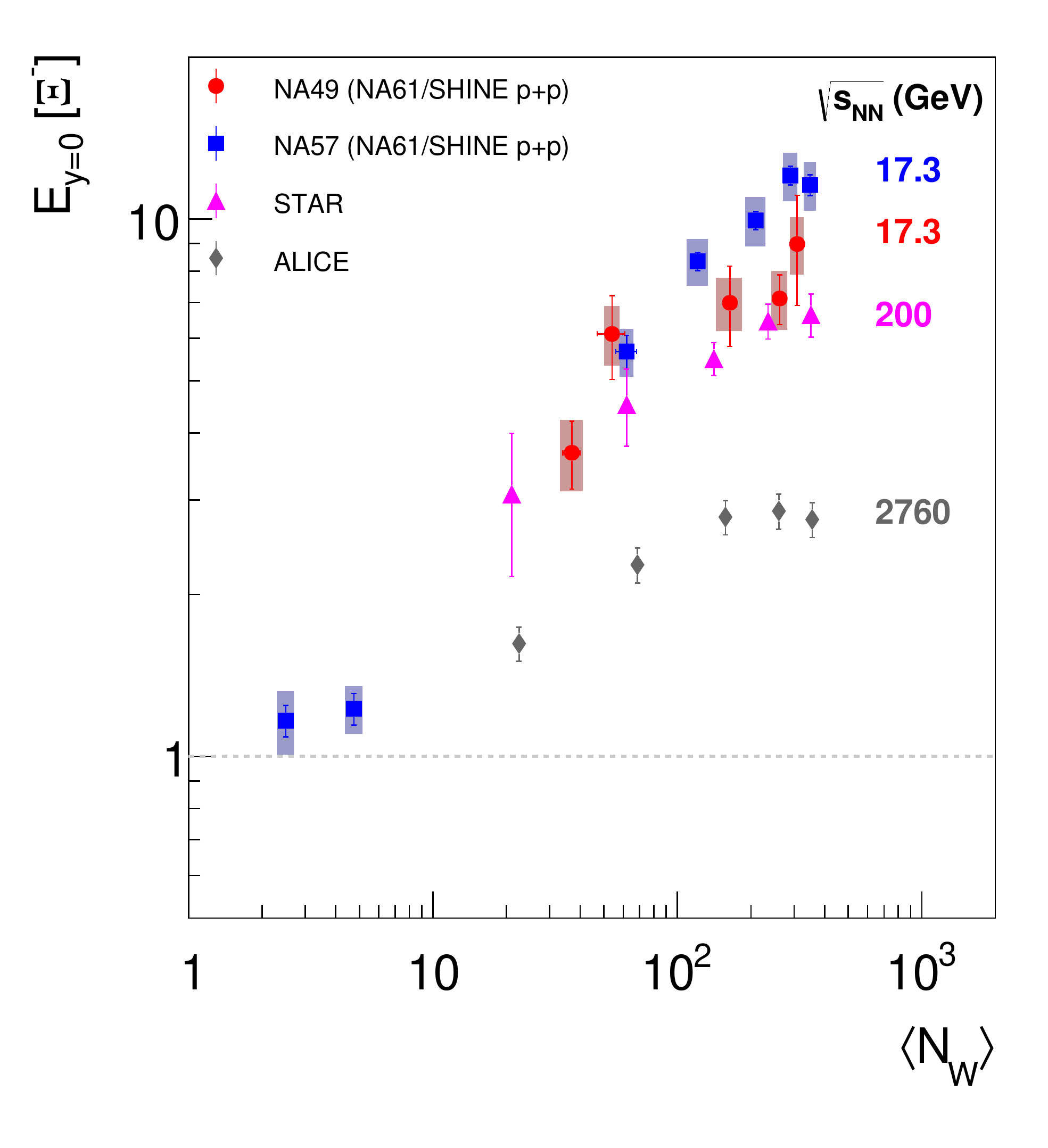}
\includegraphics[width=.49\textwidth]{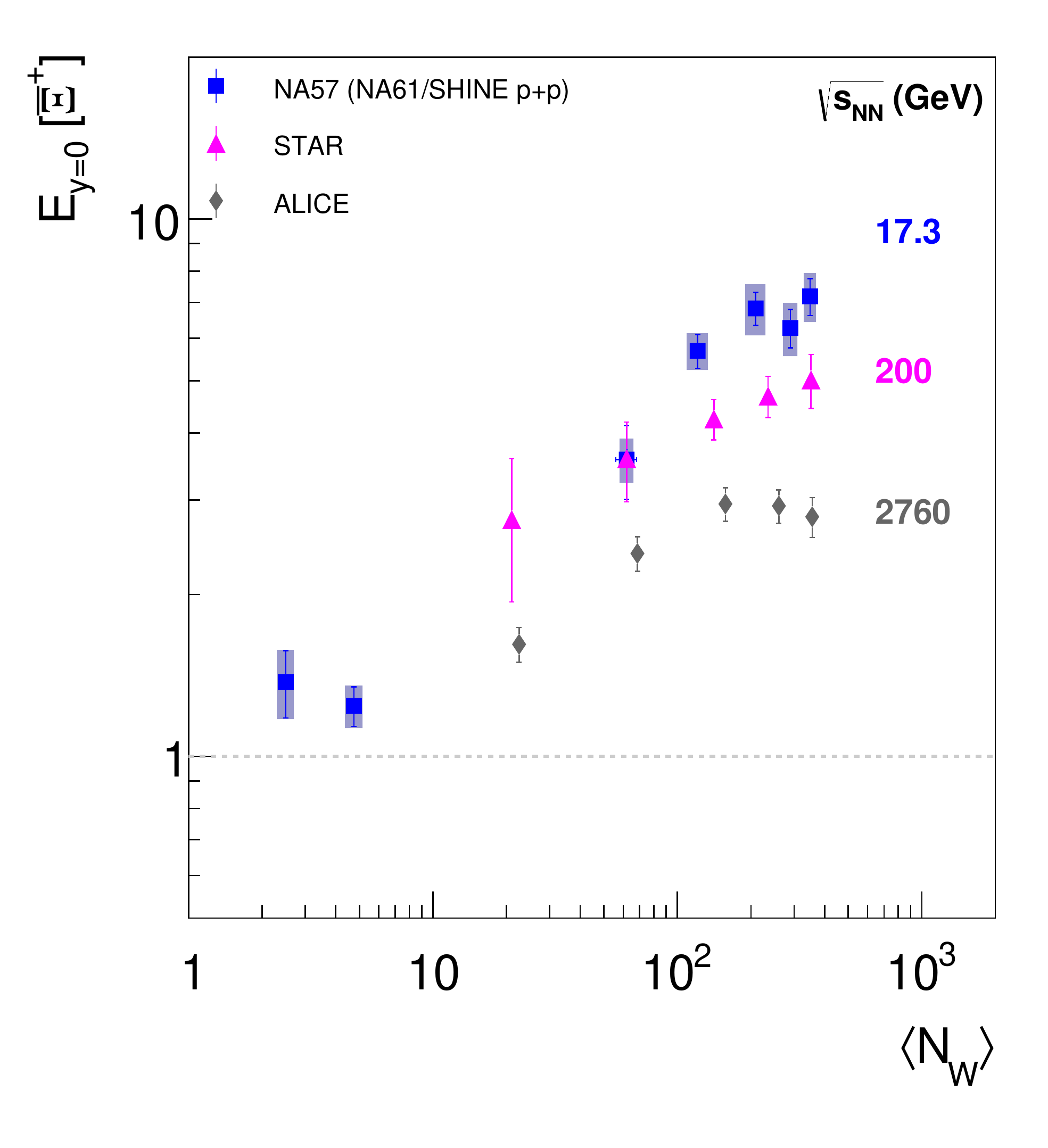}
\caption{(Color online) The strangeness enhancement $E$ at the mid-rapidity as a function of average
number of wounded nucleons  $\left<N_{W}\right>$ calculated as a ratio of rapidity density for
$\Xim$ production (\textit{left}) and $\Xip$ production (\textit{right}) in nucleus-nucleus
interactions per $\left<N_{W}\right>$ divided by the corresponding value for \pp interactions (see Eq.~\ref{eq:enhancment}).
Red circles – NA49 Pb+Pb at 158\AGeV~\cite{Anticic:2009ie}, blue squares - NA57 \textit{p}+Be, \textit{p}+Pb and Pb-Pb at the same center-of-mass energy $\sqrt{s_{NN}}$ = 17.3~\GeV~\cite{Antinori:2006ij}, magenta triangles - STAR
Au+Au at $\sqrt{s_{NN}}$ = 200~\GeV~\cite{Abelev:2007xp}, gray diamonds - ALICE Pb+Pb at $\sqrt{s_{NN}}$ = 2.76~\TeV~\cite{ABELEV:2013zaa}. The systematic errors are represented by shaded boxes.
}
\label{fig:enhancment}
\end{figure*}

The $\Xi$ mean multiplicities measured by \NASixtyOne in inelastic \pp interactions are used to calculate the enhancement factors of $\Xi$s observed in centrality selected Pb+Pb, in semi-central C+C, and in Si+Si collisions as measured by NA49 \cite{Anticic:2009ie} at the CERN SPS. The results for mid-rapidity densities are shown in Fig.~\ref{fig:enhancment} (\textit{left}) as a function of $\left<N_{W}\right>$. The enhancement factor increases approximately linearly from 3.5 in C+C to 9 in central Pb+Pb collisions. This result is compared to data from the NA57 experiment at the SPS~\cite{Antinori:2006ij}, the STAR experiment at the Relativistic Heavy Ion Collider (RHIC)~\cite{Abelev:2007xp} and the ALICE experiment at the Large Hadron Collider (LHC)~\cite{ABELEV:2013zaa}. The published enhancement factor reported by NA57 at the CERN SPS was computed using \textit{p}+Be instead of inelastic \pp interactions. Since strangeness production is already slightly enhanced in \pA collisions \cite{Susa:2002rf}, this is not a proper reference. With the advent of the \NASixtyOne results on $\Xi$ production in \pp interactions a new baseline reference becomes available and it is used here for the recalculation of the enhancement observed in the NA57 \textit{p}+Be and A+A data. The STAR Collaboration published results on multi-strange hyperon production in Au+Au collisions at $\sqrt{s_{NN}}$ from 7.7 to 39~\GeV~\cite{Adam:2019koz}, however the corresponding data on \pp and \pA interactions are missing.
The agreement between the enhancement factors calculated using the NA49 and the NA57 A+A (\textit{p}+Be) data is satisfactory. The STAR data show a slightly lower enhancement, but the enhancement observed by ALICE is significantly lower. 
Figure~\ref{fig:enhancment} (\textit{right}) shows the rapidity densities $dn/d\y$ of $\Xip$ at mid-rapidity per mean number of wounded nucleons divided by the corresponding values for inelastic \pp collisions as a function of $\left<N_{W}\right>$. Apart from a slightly flatter rise the overall picture remains unchanged.

Note that ALICE finds that the mid-rapidity yields of 
multi-strange hyperons in \pp interactions at the LHC, relative to pions, increase significantly with the charged-particle multiplicity~\cite{ALICE:2017jyt}.
\NASixtyOne results on rapidity densities of charged hadrons in \pp interactions are published~\cite{Aduszkiewicz:2017sei, Abgrall:2013qoa} allowing to include the results from the CERN SPS in the
studies of multiplicity dependence.



\section{Comparison with models}\label{sec:CompModel}

The \NASixtyOne data on charged $\Xi$ production in inelastic \pp interactions are important for the understanding of multi-strange particle production in elementary hadron interactions. In particular, the new \NASixtyOne results constitute essential input for theoretical concepts needed for the modelling of elementary hadron interactions and of more complex reactions involving nuclei like \pA and A+A collisions.

\begin{figure*}[ht!]
\centering
\includegraphics[width=.49\textwidth]{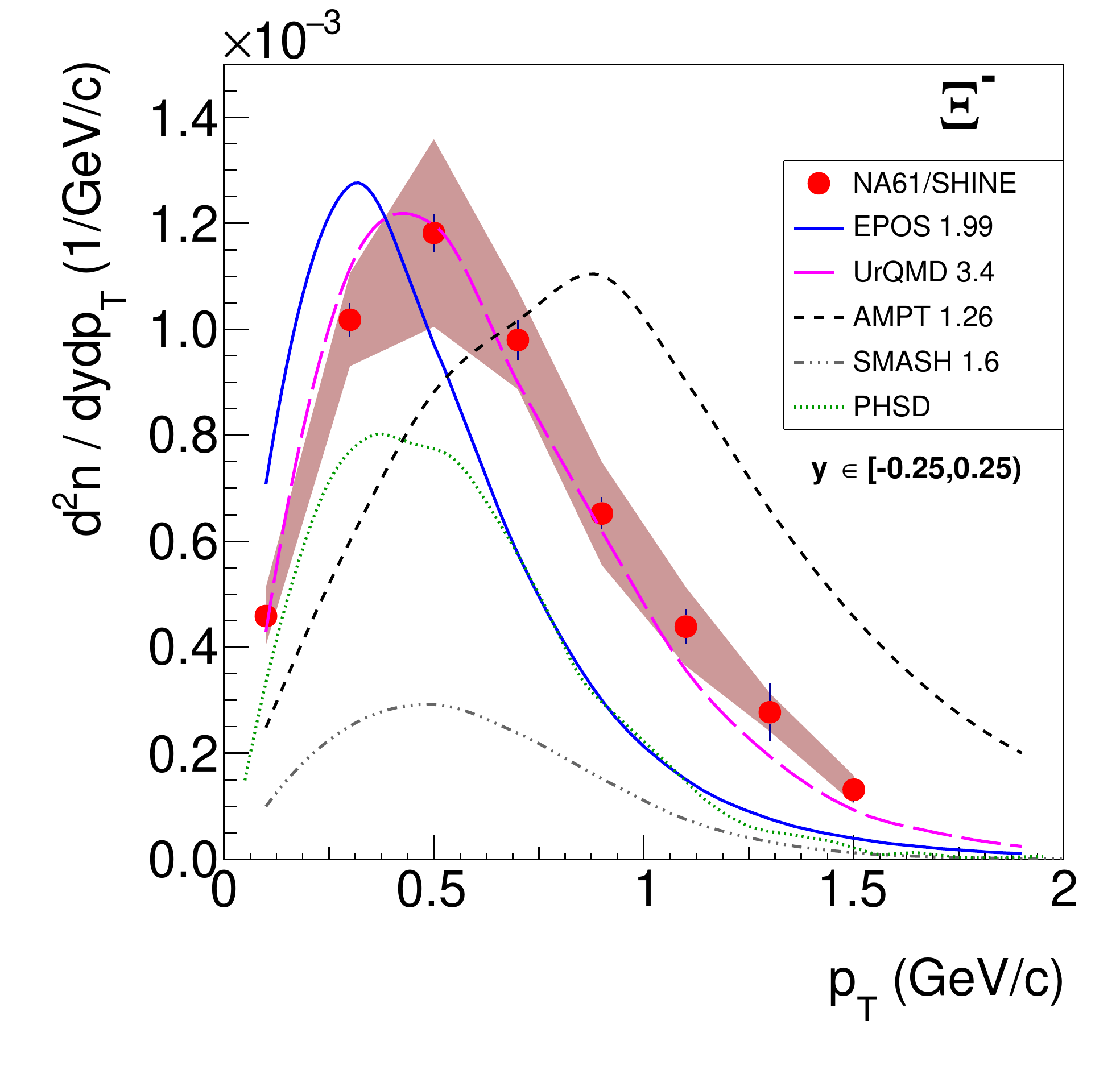}
\includegraphics[width=.49\textwidth]{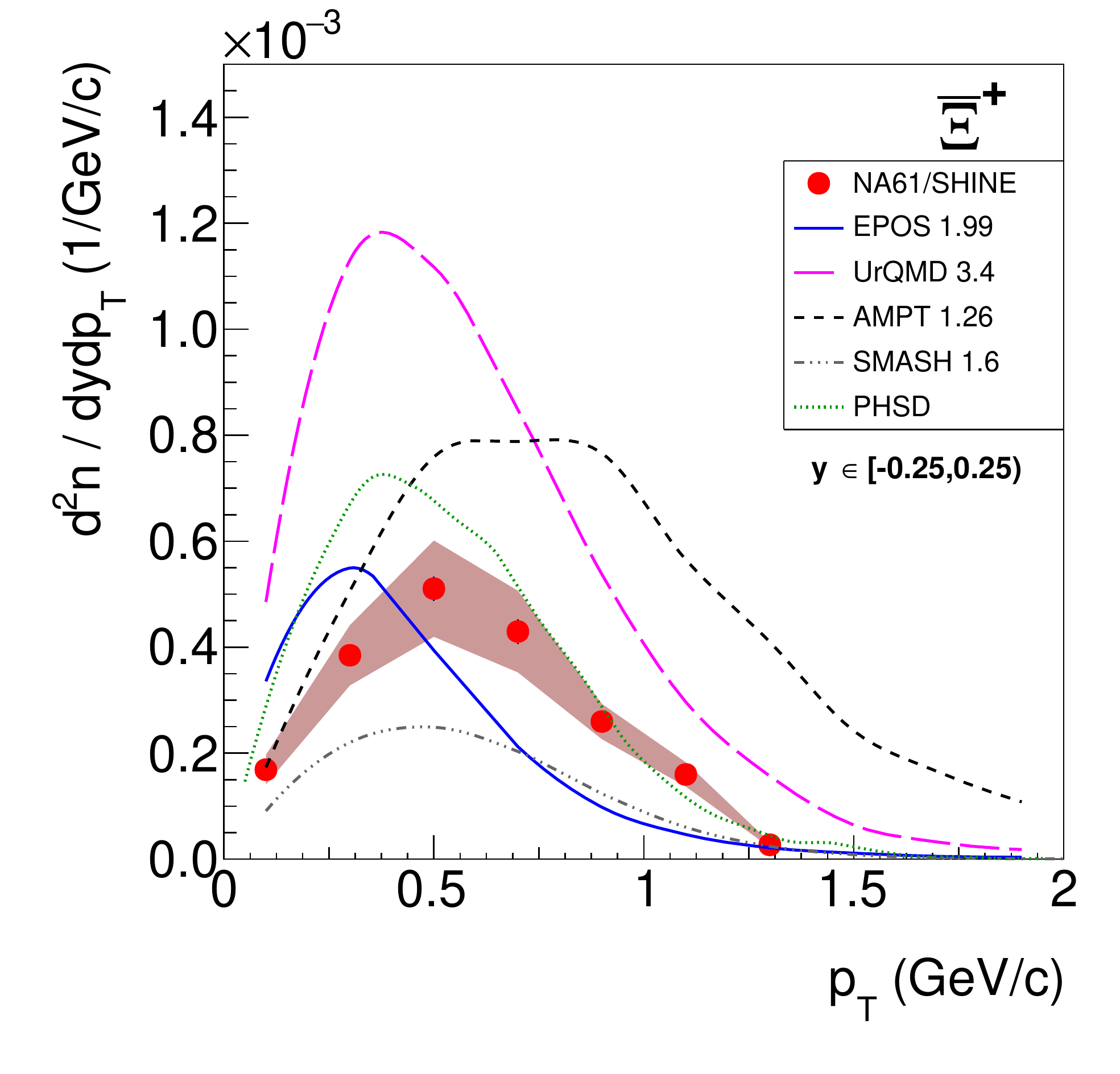}
\caption{(Color online) Transverse momentum spectra at mid-rapidity of $\Xim$ (\textit{left}) and $\Xip$ (\textit{right}) produced in inelastic \pp interactions at 158~\GeVc. Rapidity range is included in the legends. Shaded bands show systematic uncertainties. \Urqmd 3.4~\cite{Bass:1998ca,Bleicher:1999xi}, \Epos~1.99~\cite{Werner:2008zza}, \Ampt~1.26~\cite{PhysRevC.72.064901, PhysRevC.90.014904, PhysRevC.61.067901}, \SmashModel~1.6~\cite{Mohs:2019iee, Weil:2016zrk, dmytro_oliinychenko_2019_3485108} and \PHSD~\cite{Cassing:2009vt, Cassing:2008sv} predictions are shown as magenta, blue, black, gray and green lines, respectively.}
\label{fig:ptmodel}
\end{figure*}

\begin{figure*}[ht!]
\centering
\includegraphics[width=.325\textwidth]{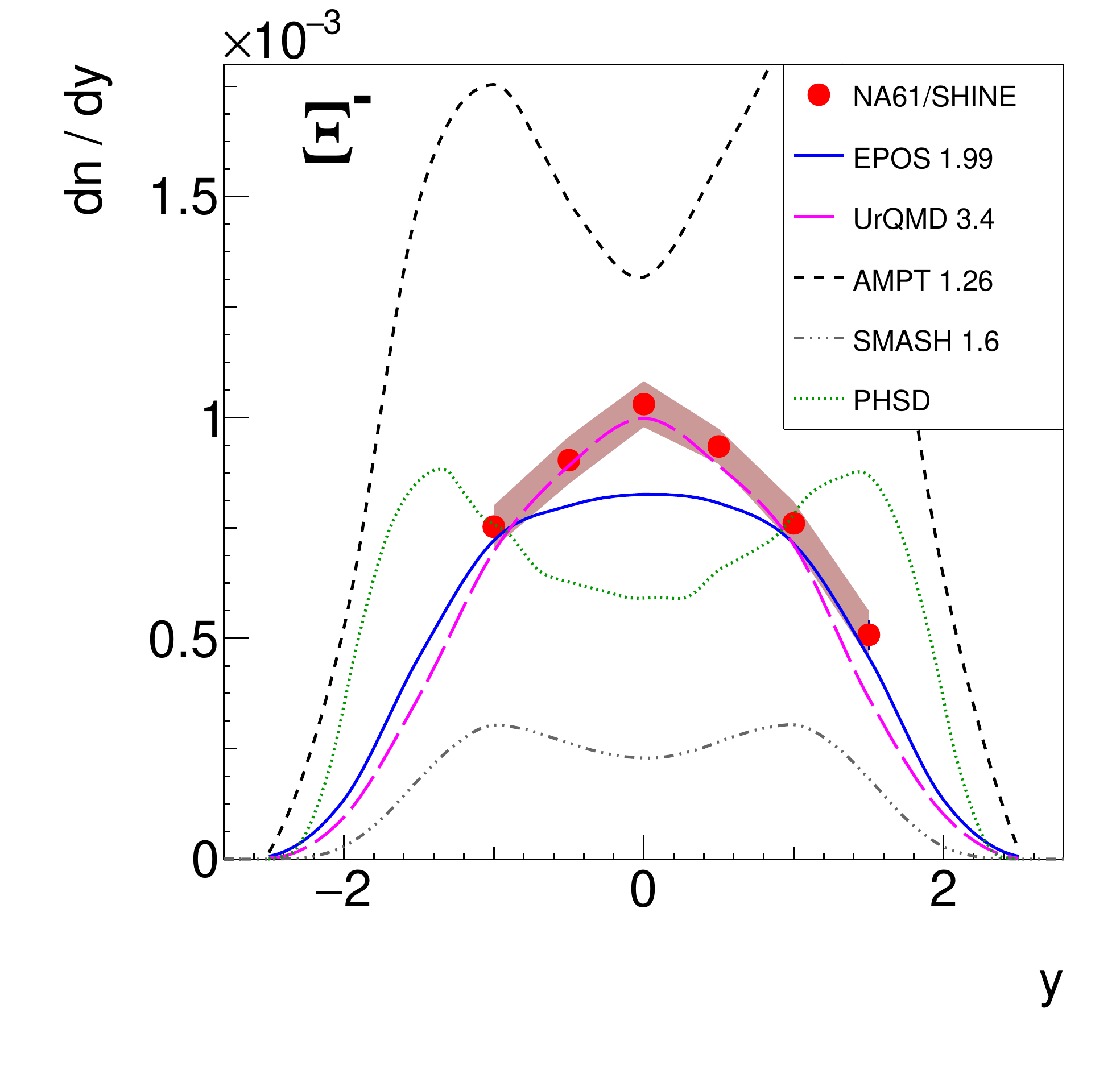}\llap{\parbox[b]{0.16\textwidth}{(a)\\\rule{0ex}{1.65in}}}
\includegraphics[width=.325\textwidth]{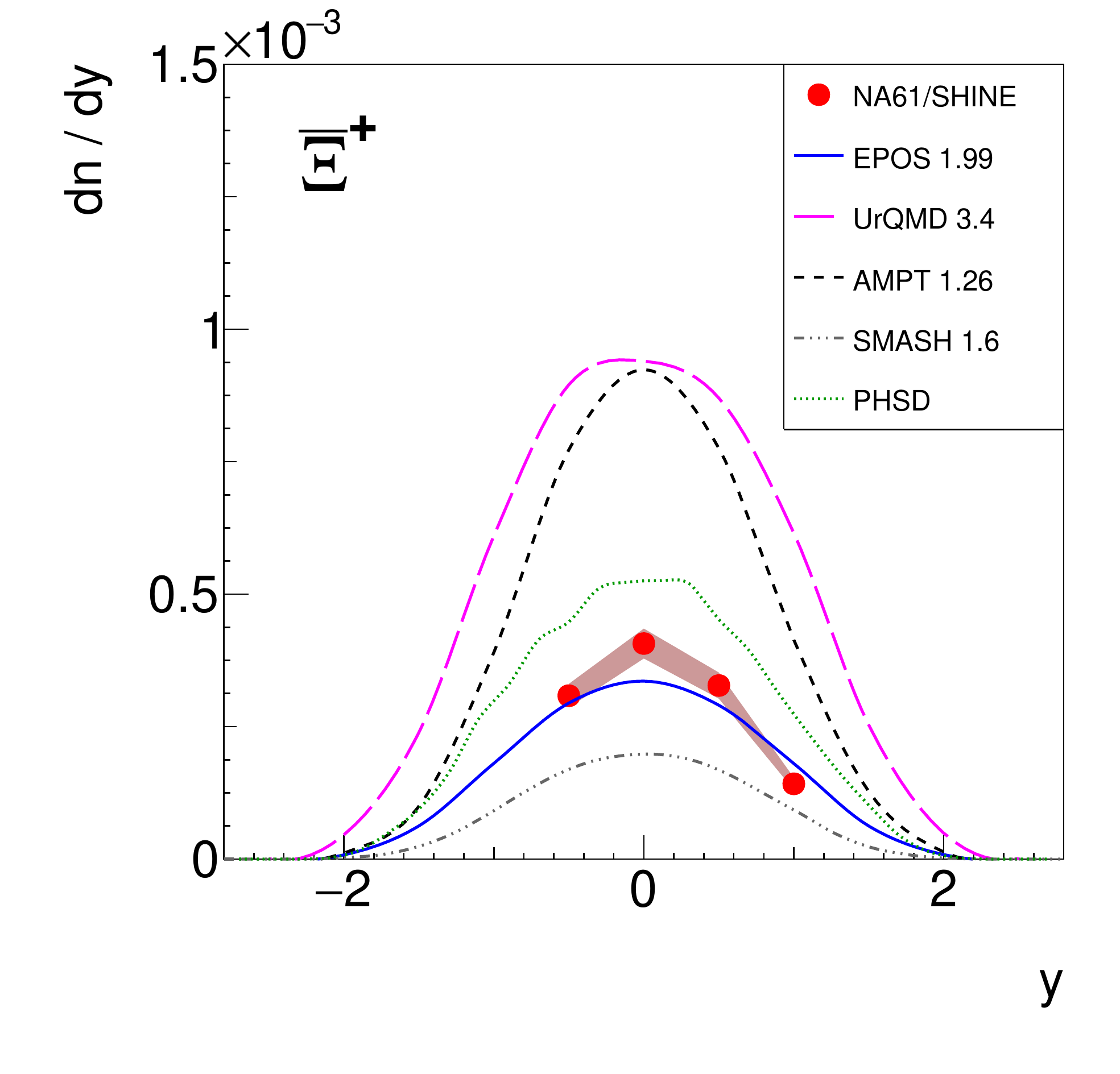}\llap{\parbox[b]{0.16\textwidth}{(b)\\\rule{0ex}{1.65in}}}
\includegraphics[width=.325\textwidth]{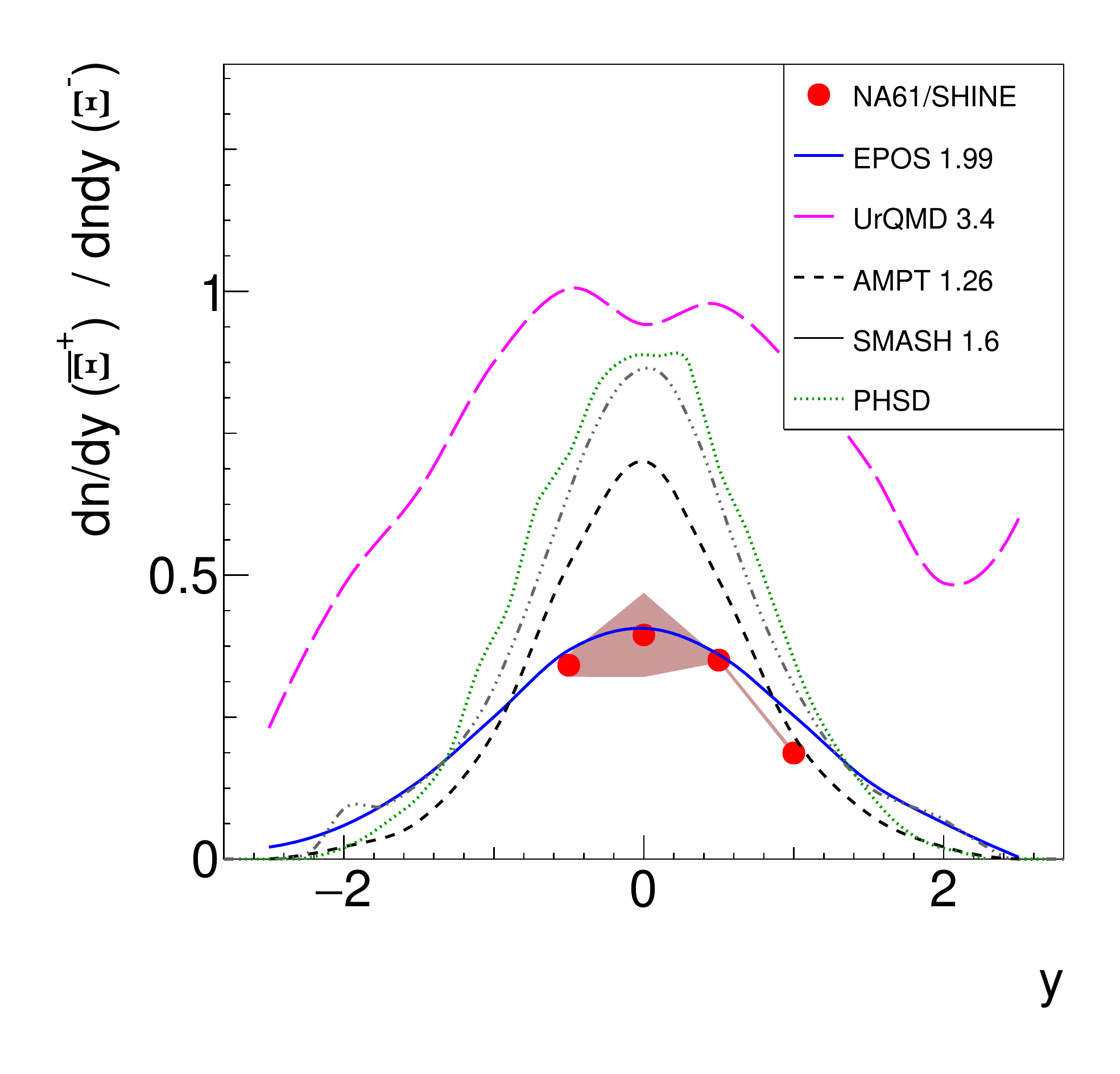}\llap{\parbox[b]{0.16\textwidth}{(c)\\\rule{0ex}{1.65in}}}
\caption{(Color online) Rapidity spectra of $\Xim$ (\textit{left}), $\Xip$ (\textit{middle}) and $\Xip/\Xim$ ratio (\textit{right}) measured in inelastic \pp interactions at 158~\GeVc. Shaded bands show systematic uncertainties. \Urqmd 3.4~\cite{Bass:1998ca,Bleicher:1999xi}, \Epos~1.99~\cite{Werner:2008zza}, \Ampt~1.26~\cite{PhysRevC.72.064901, PhysRevC.90.014904, PhysRevC.61.067901}, \SmashModel~1.6~\cite{Mohs:2019iee, Weil:2016zrk, dmytro_oliinychenko_2019_3485108} and \PHSD~\cite{Cassing:2009vt, Cassing:2008sv} predictions are shown as magenta, blue, black, gray and green lines, respectively.}
\label{fig:dnmodel}
\end{figure*}

In this section the experimental results of \NASixtyOne are compared with predictions of the following microscopic models: \Epos 1.99~\cite{Werner:2008zza}, \Urqmd 3.4~\cite{Bass:1998ca,Bleicher:1999xi}, \Ampt~1.26~\cite{PhysRevC.72.064901, PhysRevC.90.014904, PhysRevC.61.067901}, \SmashModel~1.6~\cite{Mohs:2019iee, Weil:2016zrk, dmytro_oliinychenko_2019_3485108} and \PHSD~\cite{Cassing:2009vt, Cassing:2008sv}. In \Epos the
reaction proceeds from the excitation of strings according to Gribov-Regge theory to string fragmentation 
into hadrons. \Urqmd starts with a hadron cascade on the basis of elementary cross sections for resonance production which either decay (mostly at low energies) or are converted into strings which fragment into hadrons (mostly at high energies). \Ampt uses the heavy ion jet interaction generator (\Hijing) for generating the initial conditions, Zhang's parton cascade for modeling partonic scatterings, the Lund string fragmentation model or a quark coalescence model for hadronization. \SmashModel uses the hadronic transport approach where the free parameters of the string excitation and decay are tuned to match the experimental measurements in elementary proton–proton collisions. \PHSD is a microscopic off-shell transport approach that consistently describes the full evolution of a relativistic heavy-ion collision from the initial hard scatterings and string formation through the dynamical deconfinement phase transition to the quark-gluon plasma as well as hadronization and the subsequent interactions in the hadronic phase. 
The model predictions are compared with the \NASixtyOne data in 
figs.~\ref{fig:ptmodel} and~\ref{fig:dnmodel}.
\Epos~1.99  describes well the $\Xim$ and $\Xip$ rapidity spectra but fails on the shape of the transverse momentum distribution. The comparison of the \Urqmd~3.4 calculations with the \NASixtyOne measurements reveals major discrepancies for the $\Xip$ hyperons. The model output describes almost perfectly the rapidity and transverse momentum spectra of $\Xim$ but strongly overestimates $\Xip$ yields. Consequently also the ratio of $\Xip$ to $\Xim$ cannot be described by the \Urqmd model,  see Fig.~\ref{fig:dnmodel}(c). The \Ampt, \SmashModel and \PHSD models fail in the description of both transverse momentum spectra and rapidity distributions. \Ampt overestimates the $\Xim$ and $\Xip$ multiplicities while \SmashModel underestimates them, both failing to describe the ratio. \PHSD underestimates the $\Xim$ yields and overestimates $\Xip$. Obviously \PHSD also fails to describe the ratio. \Epos  differs from the \Urqmd, \Ampt, \SmashModel and \PHSD models in its treatment of Pomeron-Pomeron interactions and of the valence quark remnants at the string ends. 

The statistical Hadron Resonance Gas Models (HGM) can be used to predict particle multiplicities in elementary and nucleus-nucleus collisions once parameters like the chemical freeze-out temperature $T_{chem}$, the baryochemical potential $\mu_B$ and strangeness saturation parameter are fixed by fits of selected mean multiplicities of hadrons. In Ref.~\cite{Becattini:2005xt} the HGM results for $\left< \Xim \right>$ and $\left< \Xip \right>$ multiplicities were calculated for two versions of the model fits. The first one, called fit B, allowed for strangeness deviation from the equilibrium introducing the free parameter $\gamma_S$. 
In the second fit, called A, the parameter $\gamma_S$ was replaced by the mean number of strange quark pairs $\langle s \bar{s} \rangle$. 
The mean  multiplicities of $\Xi$ and $\Omega$ hyperons were excluded from the fit~B and the mean multiplicity of $\phi$ meson from the fit~A. 
Table~\ref{tab:HGM} shows the HGM predictions based on the fits~A and~B together with the experimental mean multiplicities of $\Xim$ and $\Xip$ produced in inelastic \pp interactions at 158~\GeVc. 
The measurements are close to the HGM results for the fit~A which excludes mean multiplicity of $\phi$ meson. The resulting yield of $s\bar{s}$ quark pairs is about two times lower than the equilibrium one.

\begin{table}
	\centering
		\caption{The mean multiplicity of $\Xim$ and $\Xip$ hyperons produced in inelastic \pp interactions at 158~\GeVc compared to theoretical multiplicities obtained within Hadron Gas Models~\cite{Becattini:2005xt}.}
	\label{tab:HGM}
	\begin{tabular}{|c|c|c|}
	    \multicolumn{3}{c}{}\\
		\hline
		&&\\
		& $\langle \Xim \rangle$ $\times10^{-3}$ & $\langle \Xip \rangle$ $\times10^{-4}$\\
		&&\\
		\hline
		&&\\
		\NASixtyOne &  3.3 $\pm$ 0.1 $\pm$ 0.6 &   7.9 $\pm$ 0.2 $\pm$ 1.0\\
		&&\\
		\hline
		&&\\
		HGM, Canonical Ensemble, fit A (no $\phi$) \cite{Becattini:2005xt} & 2.85 & 9.18\\
		&&\\
		\hline
		&&\\
		HGM, Canonical Ensemble, fit B (with $\phi$) \cite{Becattini:2005xt} & 1.10  & 3.88 \\
		&&\\
		\hline
	\end{tabular}

\end{table}

\section{Summary}

Measurements of $\Xim$ and $\Xip$ spectra in inelastic \pp interactions at 158~\GeVc were performed by the \NASixtyOne experiment at the CERN SPS. These measurements were compared with the results obtained at higher energies, and it was shown that the mid-rapidity $\Xim$/$\Xip$ ratio in \pp at $\sqrt{s_{NN}}$ = 17.3~\GeV is around 0.5, while at higher energies it becomes unity. The \NASixtyOne results were also compared with the measurements in A+A collisions at the same energy. The ratio of rapidity densities $dn/d\y$ of $\Xim$ measured in nucleus-nucleus collisions and inelastic \pp collisions at 158\AGeV, when normalised to the same averaged number of wounded nucleons $\left<N_{W}\right>$,  rises rapidly from \pp towards peripheral Pb+Pb collisions. This strangeness enhancement was found to decrease with increasing centre-of-mass energy. Furthermore, the \NASixtyOne results were compared with \Urqmd, \Epos, \Ampt, \SmashModel and \PHSD model predictions. It was concluded that the \Epos string model provides the best description of the \NASixtyOne measurements. Finally, the mean multiplicities of $\Xim$ and $\Xip$ hyperons were compared with predictions of the Hadron Gas Model. It turned out that the HGM predictions are very close to the experimental results when the $\phi$ meson is excluded from the HGM fit.

\section*{Acknowledgments}
We would like to thank the CERN EP, BE, HSE and EN Departments for the
strong support of NA61/SHINE.

This work was supported by
the Hungarian Scientific Research Fund (grant NKFIH 123842\slash123959),
the Polish Ministry of Science
and Higher Education (grants 667\slash N-CERN\slash2010\slash0,
NN\,202\,48\,4339 and NN\,202\,23\,1837), the National Science Centre Poland (grants~2014\slash13\slash N\slash
ST2\slash02565, 2014\slash14\slash E\slash ST2\slash00018,
2014\slash15\slash B\slash ST2\slash02537 and
2015\slash18\slash M\slash ST2\slash00125, 2015\slash 19\slash N\slash ST2 \slash01689, 2016\slash23\slash B\slash ST2\slash00692,
2017\slash 25\slash N\slash ST2\slash 02575,
2018\slash 30\slash A\slash ST2\slash 00226,
2018\slash 31\slash G\slash ST2\slash 03910),
the Russian Science Foundation, grant 16-12-10176 and 17-72-20045,
the Russian Academy of Science and the
Russian Foundation for Basic Research (grants 08-02-00018, 09-02-00664
and 12-02-91503-CERN), the Ministry of Science and
Education of the Russian Federation, grant No.\ 3.3380.2017\slash4.6,
 the National Research Nuclear
University MEPhI in the framework of the Russian Academic Excellence
Project (contract No.\ 02.a03.21.0005, 27.08.2013),
the Ministry of Education, Culture, Sports,
Science and Tech\-no\-lo\-gy, Japan, Grant-in-Aid for Sci\-en\-ti\-fic
Research (grants 18071005, 19034011, 19740162, 20740160 and 20039012),
the German Research Foundation (grant GA\,1480/8-1), the
Bulgarian Nuclear Regulatory Agency and the Joint Institute for
Nuclear Research, Dubna (bilateral contract No. 4799-1-18\slash 20),
Bulgarian National Science Fund (grant DN08/11), Ministry of Education
and Science of the Republic of Serbia (grant OI171002), Swiss
Nationalfonds Foundation (grant 200020\-117913/1), ETH Research Grant
TH-01\,07-3 and the Fermi National Accelerator Laboratory (Fermilab), a U.S. Department of Energy, Office of Science, HEP User Facility managed by Fermi Research Alliance, LLC (FRA), acting under Contract No. DE-AC02-07CH11359 and the IN2P3-CNRS (France).

\bibliographystyle{na61Utphys}
\bibliography{include/na61References}
\newpage
{\Large The \NASixtyOne Collaboration}
\bigskip
\begin{sloppypar}

\noindent
A.~Aduszkiewicz$^{\,15}$,
E.V.~Andronov$^{\,21}$,
T.~Anti\'ci\'c$^{\,3}$,
V.~Babkin$^{\,19}$,
M.~Baszczyk$^{\,13}$,
S.~Bhosale$^{\,10}$,
A.~Blondel$^{\,4}$,
M.~Bogomilov$^{\,2}$,
A.~Brandin$^{\,20}$,
A.~Bravar$^{\,23}$
% ...

\end{sloppypar}

\noindent
% ...
$^{2}$~Faculty of Physics, University of Sofia, Sofia, Bulgaria\\
$^{3}$~Ru{\dj}er Bo\v{s}kovi\'c Institute, Zagreb, Croatia\\
$^{4}$~LPNHE, University of Paris VI and VII, Paris, France\\
% ...
$^{10}$~Institute of Nuclear Physics, Polish Academy of Sciences, Cracow, Poland\\
% ...
$^{13}$~AGH - University of Science and Technology, Cracow, Poland\\
% ...
$^{15}$~University of Warsaw, Warsaw, Poland\\
% ...
$^{19}$~Joint Institute for Nuclear Research, Dubna, Russia\\
$^{20}$~National Research Nuclear University (Moscow Engineering Physics Institute), Moscow, Russia\\
$^{21}$~St. Petersburg State University, St. Petersburg, Russia\\
% ...
$^{23}$~University of Geneva, Geneva, Switzerland\\
% ...

\end{document}